\documentclass{vldb}

\newfont{\mycrnotice}{ptmr8t at 7pt}
\newfont{\myconfname}{ptmri8t at 7pt}
\let\crnotice\mycrnotice%
\let\confname\myconfname%

\usepackage{latexsym}
\usepackage{newlfont}
\usepackage{algorithm}
\usepackage{algpseudocode}
\usepackage{epic}
\usepackage{eepic}
\usepackage{epsfig}
\usepackage{float}
\usepackage{graphicx}
\usepackage{caption}
\usepackage[caption=false]{subfig}
\usepackage{setspace}
\usepackage[colorlinks=true,linkcolor=blue,citecolor=blue,urlcolor=blue]{hyperref}
\usepackage{breakurl} 
\usepackage{balance}

\usepackage{caption}
\usepackage{setspace}

\algnewcommand{\LineComment}[1]{\State \(\triangleright\) #1}
\algdef{SE}{PForeach}{EndPForeach}[1]{\textbf{pforeach} \(\mbox{#1}\) \textbf{do}}{\textbf{end pforeach}}%
\algdef{SE}{Foreach}{EndForeach}[1]{\textbf{foreach} \(\mbox{#1}\) \textbf{do}}{\textbf{end foreach}}%

\algtext*{EndPForeach}
\algtext*{EndForeach}

\newcommand{\COMMENT}[1]{}

\newtheorem{defn}{Definition}[section] 

\newtheorem{prop}{Proposition}

\newcommand{\DPOP}{DynaPop}

\begin{document}

\title{Fishing in the Stream:\\ Similarity Search over Endless Data}

\numberofauthors{3}
\author{
\alignauthor
Naama Kraus\\
	\affaddr{Viterbi EE Department}\\
    \affaddr{Technion, Haifa, Israel}\\
\alignauthor
David Carmel\\
	\affaddr{Yahoo Research}
\alignauthor
Idit Keidar\\
	\affaddr{Viterbi EE Department}\\
    \affaddr{Technion, Haifa, Israel}\\
    \affaddr{and Yahoo Research}
}

\maketitle

\begin{abstract}
Similarity search is the task of retrieving data items that are similar to a given query.
In this paper, we introduce 
the time-sensitive notion of \emph{similarity search over endless data-streams (SSDS)},
which takes into account data quality and temporal characteristics in addition to similarity.
SSDS is challenging as it needs to process unbounded data,
while computation resources are bounded. 
We propose \emph{Stream-LSH}, a randomized
SSDS algorithm that bounds the index size
by retaining items according to their freshness, quality, and dynamic popularity attributes.
We analytically show that Stream-LSH increases the probability to find similar
items compared to alternative approaches using the same space capacity.
We further conduct an empirical study using real world stream datasets,
which confirms our theoretical results. 
\end{abstract}


\section{Introduction}
Users today are exposed to massive volumes of information arriving in endless data streams:  
hundreds of millions of content items are generated daily by billions of users 
through widespread social media 
platforms~\cite{Sundaram2013,Twinder2012,zephoria};
fresh news headlines from different sources around the world 
are aggregated and spread by online news services~\cite{Liu2010,Das2007}.
In this era of information explosion it has become crucial to `fish in the stream',
namely, identify stream content that will be of interest to a given user. 
Indeed, search and recommendation services that find such content are ubiquitously 
offered by major content providers~\cite{Das2007,Diaz2009,Liu2010,Chen2011,unicorn2013}.

A fundamental building block for search and recommendation applications
is \emph{similarity search},
an algorithmic primitive for finding similar content to a queried item~\cite{NNSTutorial,Andoni09NNS}.
For example, a user reading a news item or a blog post
can be offered similar items to enrich his reading experience~\cite{TwitterSearch11}.
In the context of streams, 
many works have observed that 
applications ought to take into account 
temporal metrics in addition to 
similarity~\cite{Diaz2009,Liu2010,SCENE2011,TemporalIR,SNR2014,Twinder2012,Guy:2012,Petrovic2010,Sundaram2013,SSSJ16}.
Nevertheless,
the similarity search primitive has not been extended to handle endless data-streams.
To this end, we introduce in Section \ref{sec:ssds} the problem of \emph{similarity search over data streams (SSDS)}.

In order to efficiently retrieve such content at runtime,
an SSDS algorithm needs to maintain an \emph{index} of streamed data.
The challenge, however, is that the stream is unbounded, whereas physical space capacity cannot grow without bound;
this limitation is particularly acute when the index resides in RAM for fast retrieval~\cite{Sundaram2013,Magdy16}.
A key aspect of an SSDS algorithm is therefore its \emph{retention policy},
which continuously determines which data items to retain in the index and which to forget.
The goal is to retain items that best satisfy the needs of users of stream-based applications.

In Section \ref{sec:alg}, we present \emph{Stream-LSH}, an SSDS algorithm based on \emph{Locality Sensitive Hashing (LSH)},
which is a widely used randomized similarity search technique
for massive high dimensional datasets~\cite{Gionis99}.
LSH builds a hash-based index with some redundancy in order to increase recall,
and Stream-LSH further takes into account quality, age, and dynamic popularity in determining an item's level of redundancy.

A straightforward approach for bounding the index size is to focus on the freshest items.
Thus,
when indexing an endless stream, 
one can bound the index size by
eliminating the oldest items from the index
once its size exceeds a certain threshold.
We refer to this retention policy as \emph{Threshold}.
Although such an approach has been effectively used for detecting new stories~\cite{Sundaram2013}
and streaming similarity self-join~\cite{SSSJ16},
it is less ideal for search and recommendations,
where old items are known to be valuable to users~\cite{WhatIsTwitter10,TwitterSearch11,FlickrTemporalSearch16}.

\begin{figure}[hbt]
    \centering
    \includegraphics[scale=0.4, clip]{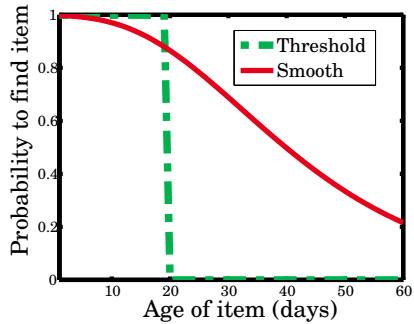}
    \vspace{1em}
    \caption{Probability of successful retrieval of similar items as a function of their age
		         with Threshold and Smooth retention policies in example settings.}    
						\label{fig:introFig}
\end{figure}

We suggest instead \emph{Smooth} -- a randomized  retention policy
that gradually eliminates index entries over time.
Since there is redundancy in the index,
items do not disappear from it at once.
Instead, an item's representation in the index decreases with its age.
Figure \ref{fig:introFig} illustrates the probability to find a similar item with the two
retention policies using the same space capacity.
In this example, the index size suffices for Threshold
to retain items for $20$ days.
We see that Threshold is likely to find fresh similar items,
but fails to find items older than $20$.
Using the same space capacity, 
Smooth finds similar items for a longer time period
with a gradually decaying probability;
this comes at the cost of a lower probability to find very fresh items.
We further show that
Smooth exploits capacity resources more efficiently so that the average recall 
is larger than with Threshold.

We extend Stream-LSH to consider additional data characteristics beyond age.
First, our Stream-LSH algorithm considers items' query-independent quality,
and adjusts an item's redundancy in the index based on its quality.
This is in contrast to the standard LSH, 
which indexes the same number of copies for all items regardless of their quality.
Second, we present the \emph{\DPOP{}} extension to Stream-LSH, 
which considers items' dynamic popularity.
\DPOP{} gets as input a stream of user interests in items, such as retweets or clickthrough information, 
and re-indexes in Stream-LSH items of interest; 
thus, it has Stream-LSH dynamically adjust items' redundancy to reflect their popularity.

To analyze Stream-LSH with different retention policies, 
we formulate in Section \ref{sec:analysis} the theoretical \emph{success probability (SP)} 
metric of an SSDS algorithm when seeking
items within given similarity, age, quality, and popularity radii.
Our results show that Smooth increases the probability to find similar and high quality items
compared to Threshold, when using the same space capacity.
We show that our quality-sensitive approach 
is appealing for similarity search applications that handle large amounts 
of low quality data, such as user-generated social data~\cite{Agichtein2008,Becker11,EarlyBird},
since it increases the probability to find high-quality items.
Finally, we show that using \DPOP{},
Stream-LSH is likely to find popular items that are similar to the query,
while also retrieving similar items that are not highly popular albeit with lower probability.
Retrieving similar items from the tail of the popularity distribution in addition to the most popular ones 
is beneficial for applications such as query auto-completion~\cite{BarYossef11} and product recommendation~\cite{LongTailRec12}.
In Section \ref{sec:eval}
we validate our theoretical results empirically on several real-world stream datasets using the recall metric.

In summary, we make the following contributions:
\begin{itemize}
 \item We formulate SSDS -- a time-sensitive similarity search primitive for unbounded data streams
	(Section \ref{sec:ssds}). 
\item We propose Stream-LSH, 
      a bounded-capacity SSDS algorithm with randomized insertion and retention policies (Section \ref{sec:alg}).
\item We show both analytically and empirically that Stream-LSH successfully finds similar
items according to age, quality, and dynamic popularity attributes (Sections \ref{sec:analysis} and \ref{sec:eval}).
\end{itemize}

Related work is discussed in Section \ref{sec:RelatedWork},
and Section \ref{sec:conc} concludes the paper.

\section{Similarity Search over Data-Streams}
\label{sec:ssds}
We extend the problem of similarity search
and define similarity search over unbounded data-streams.
Our stream similarity search is time-sensitive and quality-aware,
and so we also define a recall metric that takes these aspects into account.

\subsection{Background: Similarity Search}
\label{sec:prelm_SS}
Similarity search is based on a \emph{similarity function},
which measures the similarity between two vectors, 
$u,v\in V$, where $V=(\mathbb{R}^+_0)^d$ is some high $d$-dimensional vector space~\cite{Chierichetti12}:
\begin{defn}[similarity function]
A similarity function 
$sim: V \times V \rightarrow [0,1]$
is a function
such that $\forall u,v \in V, sim(u,v) = sim(v,u)\ and \ sim(v,v)=1$.
\end{defn}
The similarity function returns a \emph{similarity value} within the range $[0,1]$,
where $1$ denotes perfect similarity, and $0$ denotes no similarity.
We say that $u$ is \emph{$s$-similar} to $v$ if $sim(u,v)=s$.

A commonly used similarity function for textual data
is \emph{angular similarity}~\cite{Sundaram2013,Petrovic2010},
which is closely related to cosine similarity~\cite{Charikar02,Chierichetti12}.
The angular similarity between two vectors $u,v \in V$ is defined as:
\begin{equation}
\label{eq:ang}
sim(u,v) = 1-\frac{\theta(u,v)}{\pi},
\end{equation}
where $\theta(u,v)=arccos(\frac{u \cdot v}{\left\|u\right\| \cdot \left\|v\right\|})$ 
is the angle between $u$ and $v$.

Given a (finite) subset of vectors, $U \subseteq V$, 
similarity search is the task of finding vectors that are similar to some query vector. 
More formally, an \emph{exact similarity search} algorithm accepts 
as input a \emph{query} vector $q \in V$ and a \emph{similarity radius} $R \in [0,1]$, 
and returns \emph{Ideal}$(q,R)$, a unique \emph{ideal result set} of all vectors $v \in U$ satisfying $sim(q,v) > R$. 

The time complexity of exact similarity search has been shown to be linear in the number of items searched,
for high-dimensional spaces~\cite{Weber1998}.
\emph{Approximate similarity search} improves search time complexity
by trading off efficiency for accuracy~\cite{NNSTutorial,Andoni09NNS}.
Given a query $q$, it returns \emph{Appx}$(q,R)$, an \emph{approximate result set} of vectors, 
which is a subset of $q$'s $R$-ideal result set.
We refer to approximate similarity search simply as similarity search in this paper.

\subsection{SSDS}
\label{sec:streamSSProb}
SSDS considers an unbounded \emph{item stream} $U \subseteq V$ arriving over an infinite time period,
divided into discrete \emph{ticks}.
The (finite) time unit represented by a tick is specified by the application,
e.g., $30$ minutes or 1 day.
On every time tick, $0$ or more new items arrive in the stream,
and the \emph{age} of a stream item is the number of time units that elapsed since its arrival.
Note that each item in $U$ appears only once at the time it is created.
Each item is associated with a query-independent quality score,
which is specified by a given weighting function $quality:V \rightarrow [0,1]$.

\paragraph*{Similarity search over data-streams}
An SSDS algorithm's input consists of a \emph{query} vector $q \in V$
and a three-dimensional radius, $(R_{sim},R_{age},R_{quality})$, 
of similarity, age, and quality radii, respectively.
An \emph{exact SSDS algorithm} returns a unique ideal result set
\begin{equation}
\begin{array}{l}
\mbox{\emph{Ideal}}(q,R_{sim},R_{age},R_{quality}) \triangleq \\ 
\ \ \ \{v \in U | sim(q,v) \ge R_{sim} \wedge age(v) \le R_{age}  \wedge \nonumber  \\ 
\ \ \  quality(v) \ge R_{quality} \}.
\end{array}
\end{equation}
An (approximate) \emph{SSDS algorithm} $A$ returns\\ 
a subset \mbox{\emph{Appx}}$(A,q,R_{sim},R_{age},R_{quality})$
of $q$'s ideal result set.

\paragraph*{Recall}
\label{sec:Recall}
\begin{defn}[recall at radius]
The \emph{recall at radius} of algorithm $A$ for query $q$ and radius $(R_{sim},R_{age},R_{quality})$ is
\[
\mbox{\emph{Recall}}(A,R_{sim},R_{age},R_{quality})(q) \triangleq
\]
\[
\frac{|Appx(A,q,R_{sim},R_{age},R_{quality})|}{|Ideal(q,R_{sim},R_{age},R_{quality})|}.
\]
The recall at radius $\mbox{\emph{Recall}}(A,R_{sim},R_{age},R_{quality})$ of $A$ is the mean recall over the query set $Q$.
\end{defn}

\paragraph*{Dynamic popularity}
We consider a second unbounded stream $I$ which consists of items from the item stream $U$
and arrives in parallel to $U$.  
We call $I$ the \emph{interest stream}.
The arrival of an item at some time tick in $I$ signals interest in the item at that point in time. 
Note that an item may appear multiple times in the interest stream.

We capture an item's \emph{dynamic popularity} by a weighted aggregation of the number of times it appears in the interest stream,
where weights decay exponentially with time~\cite{MiningMassive}:
Let $t_0,\ldots,t_n$ denote time ticks since the starting time $t_0$,
and the current time $t_n$.
The indicator $a_i(x)$ is $1$ if item $x$ appears in the interest stream at time $t_i$ and is $0$ otherwise.
A parameter $0 < \alpha < 1$ denotes the \emph{interest decay}, 
which controls the weight of the interest history and is common to all items.
\begin{defn}[item popularity]
\label{def:popScore}
The function $pop:U \rightarrow [0,1]$ assigns a popularity score pop(x) to an item $x \in U$: \\ 
\[
pop(x) \triangleq (1-\alpha) \sum_{i=0}^n a_i(x) \alpha^{(n-i)}.
\]
\end{defn}

Given an assignment of popularity scores to items,
we are interested in the retrieval of items within a popularity radius $R_{pop} \in [0,1]$,
i.e., with a popularity score that is not lower than $R_{pop}$.
We define recall in a similar manner to the previous definitions.

\section{Stream-LSH}
\label{sec:alg}
Stream-LSH is an extension of Locality Sensitive Hashing (overviewed in Section \ref{sec:lsh})
for unbounded data-streams,
augmented with age, quality, and dynamic popularity dimensions.
Stream-LSH consists of a retention policy that defines 
which items are retained in the index and which are eliminated
as new items arrive.

\subsection{Background: Locality Sensitive Hashing}
\label{sec:lsh}
Locality Sensitive Hashing (LSH)~\cite{LSH_Indyk1998,Gionis99} is a widely used approximate similarity search algorithm 
for high-dimensional spaces, with sub-linear search time complexity.
LSH limits the search to vectors that are likely to be similar to the query vector
instead of linearly searching over all the vectors.
This reduces the search time complexity at the cost of missing similar vectors with some probability.

LSH uses hash functions that map a vector in the high dimensional input space $(\mathbb{R}^+_0)^d$
into a representation in a lower dimension $k << d$,
so that the hashes of similar vectors are likely to collide.
LSH executes a pre-processing (index building) stage, 
where it assigns vectors into buckets according to their hash values.
Then, given a query vector, the similarity search algorithm computes its hashes and searches vectors in the corresponding buckets. 
The LSH algorithm is parametrized by $k$ and $L$, where $k$ is the hashed domain's dimension,
and $L$ is the number of hash functions used, as explained below.
Formally~\cite{Charikar02}: a \emph{locality sensitive hashing} with similarity function $sim$ 
is a distribution on a family $\mathcal{H}$ of hash functions on a
collection of vectors, 
$h: V \rightarrow \left\{0,1\right\}$, such
that for two vectors $u$, $v$,
\begin{equation}
Pr_{h \in \mathcal{H}}\left[h(u)=h(v)\right]=sim(u,v).
\end{equation}
We use here a hash family $\mathcal{H}$ for angular similarity~\cite{Charikar02}.
In order to increase the probability that similar vectors are mapped to the same bucket,
the algorithm defines a family $\mathcal{G}$ of hash functions,
where each $g(v) \in \mathcal{G}$ is a concatenation of $k$ functions chosen randomly and independently from $\mathcal{H}$.
In the case of angular similarity, $g: V \rightarrow \left\{0,1\right\}^k$,
i.e., $g$ hashes $v$ into a binary \emph{sketch vector}, which encodes $v$ in a lower dimension $k$.
For two vectors $u,v$, $Pr_{g \in \mathcal{G}}\left[g(u)=g(v)\right]=(sim(u,v))^k$,
for any randomly selected $g \in \mathcal{G}$.
The larger $k$ is, the higher the precision. 

In order to mitigate the probability to miss similar items, 
the $LSH$ algorithm selects $L$ functions randomly and independently from $\mathcal{G}$.
The item vectors are now replicated in $L$ hash tables $H_i$, $1 \le i \le L$.
Upon query, search is performed in $L$ buckets.
This increases the recall at the cost of additional storage and processing.


\COMMENT {
\begin{figure}[hbt]
    \centering
    \includegraphics[scale=0.3, clip]{figures/lsh_illustration.jpg}
    \caption{LSH hash table: a hash function maps a vector in a high dimensional space into a $k$-dimensional hash,
		         such that similar vectors are likely to collide.
						 One of $L$ hash tables is shown.
						 Given a query $q$, LSH searches in $q$'s bucket for items similar to it.}
						\label{fig:lsh}
\end{figure}
}

\subsection{Stream-LSH}
\label{sec:algIndexing}
Stream-LSH,
presented in Algorithm \ref{alg:stream-lsh},
extends LSH's indexing procedure to operate on an unbounded data-stream.
Every time tick, Stream-LSH accepts a set of newly arriving items in the item stream $U$
and indexes each item into its LSH buckets.
Stream-LSH selects an item's initial redundancy according to its quality:
it indexes the item into each bucket with a probability 
that equals its quality, independently of other buckets.
In addition, in order to bound the index size, in each time tick,
Stream-LSH eliminates items from the index according to the retention policy it uses.
Note that the two operations -- indexing new items and eliminating old ones -- are independent,
and indexing and elimination work independently in each bucket.

\begin{algorithm}
\caption{Stream-LSH}
\label{alg:stream-lsh}
\begin{algorithmic}[1]
\State \textbf{On every time tick $t$ do}:
\Foreach{$H_i \in HashTables$} 
  \Foreach{$item \in items(t)$} 
	\label{line:arrival}
	\LineComment{Hash to bucket $B_i$}
		\State $B_i \gets g_i(item)$ 
    \LineComment{Quality-based indexing}
		\State \mbox{with probability} $quality(item)$, $B_i.\Call{add}{item}$ 
		\label{line:index}
     \LineComment{Elimination by retention policy}
	\State $H_i.\Call{Eliminate}$
	\EndForeach	
\EndForeach
\Statex
\end{algorithmic}
\end{algorithm}

\subsection{Retention Policies}
\label{sec:retention}
We first describe the Threshold~\cite{Sundaram2013,SSSJ16} 
and Bucket~\cite{Petrovic2010} policies,
which have been used by prior art in other contexts.
Then, 
we describe our randomized Smooth policy
that gradually eliminates item's copies from the index as a function of its age.

\subsubsection{Threshold and Bucket}
\label{sec:ageBased}
The Threshold retention policy~\cite{Sundaram2013,SSSJ16}
presented in Algorithm \ref{alg:threshold}
sets a limit $T_{size}$ on table size,
and eliminates the oldest items from all $L$ tables once the size limit is exceeded.
Note that with Threshold, the number of copies of an item in the index does not vary with age.

\begin{algorithm}
\caption{Threshold retention policy}
\label{alg:threshold}
\begin{algorithmic}[1]
\Function{H.Eliminate}{}
	\State \mbox{remove $\left|H\right|$ - $T_{size}$ oldest items in table}
\EndFunction
\Statex
\end{algorithmic}
\end{algorithm}

The Bucket retention policy~\cite{Petrovic2010}
given in Algorithm \ref{alg:bucket}
sets a limit $B_{size}$ on bucket size (rather than on table size),
and eliminates the oldest items in each bucket once its size limit is exceeded.
Note that with Bucket, the number of copies of an item in the index varies with age,
since each bucket is maintained independently.
The probability of an item to be eliminated from a bucket depends on the data distribution,
i.e., on the probability that newly arriving items will be mapped to that item's bucket.

\begin{algorithm}
\caption{Bucket retention policy}
\label{alg:bucket}
\begin{algorithmic}[1]
\Function{H.Eliminate}{}
  \Foreach{$B_i \in H$}
	\State \mbox{remove $\left|B_i\right|$ - $B_{size}$ oldest items in bucket}
  \EndForeach
\EndFunction
\Statex
\end{algorithmic}
\end{algorithm}

\subsubsection{Smooth}
\label{sec:mdBased}
In Algorithm \ref{alg:smooth} we present \emph{Smooth}, 
our randomized retention policy that gradually eliminates item copies from 
the index as a function of their age.
Smooth accepts as a parameter a \emph{retention factor} $p$, $0 < p <1$.
Upon a time tick, Smooth eliminates each existing item copy from its bucket with probability $1-p$,
independently of the elimination of other items.
The number of buckets an item is indexed into thus exponentially decays over time.
As we show in Section \ref{sec:indexSizeAnalysis},
Smooth entails an expected bounded index size that is a function of $p$.

\begin{algorithm}
\caption{Smooth retention policy}
\label{alg:smooth}
\begin{algorithmic}[1]
\Function{H.Eliminate}{} 
	\Foreach{item in $H$}
		\State with probability $1-p$, remove item from $H$
	\EndForeach
\EndFunction
\Statex
\end{algorithmic}
\end{algorithm}

\COMMENT {
Note that we assume arbitrary queries therefore all item copies are eliminated with the same probability $p$. 
However, the retention policy can be easily extended if the query distribution is given.
This is done by associating a retention factor $p$ for each item based on the query distribution, 
i.e items that are more likely to be similar to the expected query are kept longer in the index (with higher $p$).
}

Although as described in Algorithm \ref{alg:smooth}
Smooth entails a linear scan of all items in all hash tables at each time unit,
it can be implemented efficiently by 
randomly and uniformly selecting a fraction of $1-p$  of the 
items to eliminate from each table.

\subsection{Dynamic Popularity}
\label{sec:dynpop}
\DPOP{} extends Stream-LSH indexing procedure to dynamically re-index items based on signals of user interests, 
as reflected by the interest stream $I$.
Here, an item's redundancy increases as the interest in it increases.
At each time tick, \DPOP{} re-indexes an item that arrives in $I$ into each of its buckets with probability $quality(x)u$ independently of other buckets;
the \emph{insertion factor}, $0 < u < 1$, is a parameter to the algorithm.
Note that in this context, an item's quality may also change dynamically over time.
At each time tick, the current quality value is considered.

\section{Analysis}
\label{sec:analysis}
In this section, we analyze Stream-LSH's index size
and prove that it maintains a bounded index in expectation.
We additionally analyze the probability that Stream-LSH finds 
items within similarity, age, quality, and popularity radii.

\subsection{Index Size and Number of Retained Copies}
\label{sec:indexSizeAnalysis}
\paragraph*{Index size}
We analyze the index size according to Stream-LSH's retention policies.
For the sake of the analysis, we assume that a constant number of new items $\mu$ arrive at each time unit,
and that their mean quality is $\phi$.
By definition, Threshold and Bucket guarantee a bounded index size.
We analyze the index size
while using Smooth with a retention factor $p$.
Consider one specific hash table,
and denote time ticks as $t_0, \cdots, t_n$.
At time $t_0$, Smooth stores $\mu \phi$ items in the hash table in expectation.
This set of $\mu \phi$ items are scanned independently $n$ times until time $t_n$,
and thus the expected number of items that arrive at $t_0$ and survive elimination until $t_n$ is $p^n\mu \phi$.
It follows that the expected number of items in the table when processing an infinite stream is
$\sum_{i=0}^{\infty}p^i\mu \phi = \frac{\mu \phi}{1-p}$.
The retention process is performed independently in each of the $L$ hash tables, therefore,
\begin{prop}
\label{prop:plshTableSize}
Assuming $\mu$ new items with mean quality $\phi$ arrive at each time unit,
the expected size of an index with $L$ hash tables when using Smooth with retention factor $p$
is $\frac{\mu \phi}{1-p}L$.
\end{prop}

\paragraph*{Number of retained copies}
Next, we analyze the evolution of an item's number of copies in the index as a function of its age and quality
according to Threshold and Smooth.
We omit Bucket from our analysis due to its dependency on the data distribution.
We examine Threshold and Smooth when using the same index size: $20\mu \phi L$ in expectation,
and so we set $T_{size} = 20\mu \phi$ for Threshold, $p=0.95$ for Smooth (Proposition \ref{prop:plshTableSize}),
and $L=15$ for both. 

\begin{figure}[hbt]
    \centering
    \includegraphics[scale=0.4, clip]{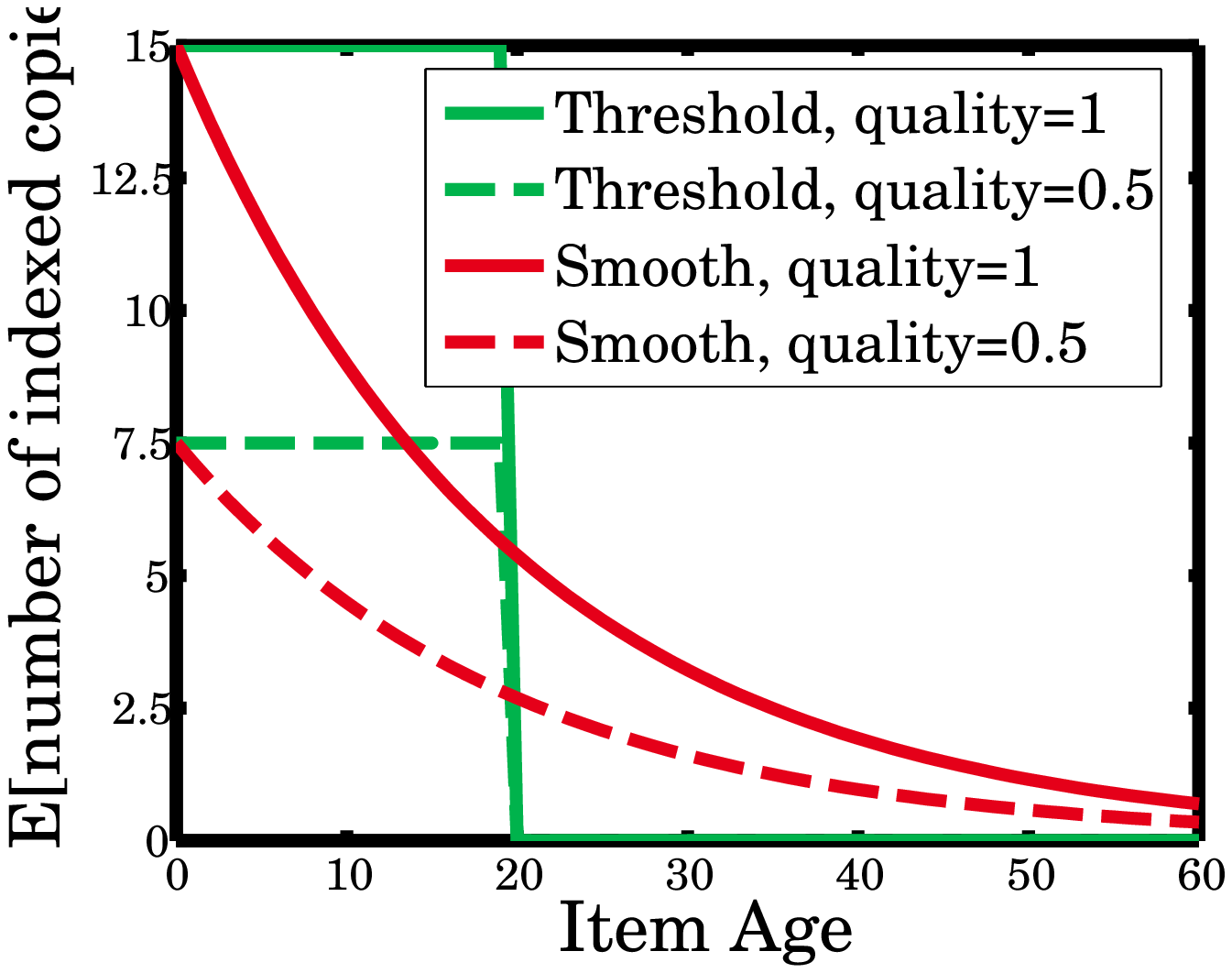}
    \vspace{1em}
    \caption{Expected number of copies of an item in the index as a function of its age, for items with quality values of $1$ and $0.5$.}
						\label{fig:nCopies}
\end{figure}

Let $x$ be some item.
Threshold retains $quality(x)L$ copies of $x$ in expectation if $age(x) < 20$,
and zero copies otherwise.
Smooth retains $quality(x)p^{age(x)}L$ copies of $x$ in expectation.
Figure \ref{fig:nCopies} illustrates the number of index copies of an item with quality $1$ (solid line),
and of an item with quality $0.5$ (dashed line), as a function of the item's age.
Due to its quality-based indexing, Stream-LSH (for both Threshold and Smooth)
retains a smaller number of copies of low quality items compared to high quality ones.
Additionally, as Smooth's name suggests, it smoothly decays an item's number of copies in contrast with Threshold.
This difference between the two policies impacts their effectiveness as we analyze in Section \ref{sec:spAnalysis}.

\subsection{Success Probability}
\label{sec:spAnalysis}
Success probability quantifies the probability of an 
algorithm to find some item $x$ given a query $q$~\cite{Multi-probe07}.
For an LSH algorithm $A$ with parameters $k$ and $L$,
we denote this probability by $SP(A(k,L))$.
We sometimes omit $k,L$ where obvious from  the context.

According to LSH theory for angular similarity~\cite{Charikar02},
the probability of an $LSH(k,L)$ algorithm 
to find $x$ that is $s$-similar to $q$ in bucket $g_i(q)$ equals $s^k$, 
under a random selection of $g_i \in \mathcal{G}$.
LSH searches in $L$ buckets $g_1(q), \cdots, g_L(q)$
independently,
thus, $LSH(k,L)$ finds $x$ with probability
$1-\left(1-s^k\right)^L$. 

\subsubsection{Retention Policies}
\label{sec:sp_rPolicies}
We analyze Stream-LSH success probability when using Threshold and Smooth.
We do not analyze Bucket, as its behavior depends on the data distribution,
which we do not model.

\paragraph*{Success probability}
We denote by $SP(A(k,L), s, a, z)$ 
the probability of algorithm $A(k,L)$ to find an item $x$ for query $q$,
s.t. $sim(q,x)=s$, $age(x)=a$,
and $quality(x) = z$.

Stream-LSH indexes a newly arriving item $x$ into each bucket $g_i(x)$ independently
with probability $z$.
For a constant arrival rate $\mu$, a mean quality $\phi$, and a size limit $T_{size}$,
Threshold eliminates items that reach age $T_{age}=\frac{T_{size}}{\mu \phi}$. 
Thus,
\begin{equation}
\label{eq:sp_agelsh}
SP(\mbox{Threshold}(k,L), s, a, z) = \left\{
  \begin{array}{@{}ll@{}}
    1-(1-s^kz)^L, & \text{if}\ a < T_{age} \\
    0, & \text{otherwise}
  \end{array}\right.	
\end{equation}

Smooth retains an item $x$ in the index with probability $p^{age(x)}$,
thus, 
\begin{equation}
\label{eq:sp_mdlsh}
SP(Smooth(k,L), s, a, z) = 1-(1-p^as^kz)^L.
\end{equation}

\paragraph*{Numerical illustration}
\label{sec:spAgeSim}
We compare the success probabilities of Threshold and Smooth.
In order to achieve a fair comparison, 
we fix $k$, $L$, and the index size.
Given that our treatment of quality is orthogonal to the retention policies,
our example ignores quality, and so we assume $quality(x)=1$ for all items $x$.
For the purpose of the illustration, we select a configuration where $k=10$ and $L=15$; 
we set $T_{size}=20\mu$ and $p=0.95$ yielding a common index size for both policies
(Proposition \ref{prop:plshTableSize}).

\begin{figure}[tbh]
        \centering
        \subfloat[SP Threshold(10,15)]{\includegraphics[scale=0.3]{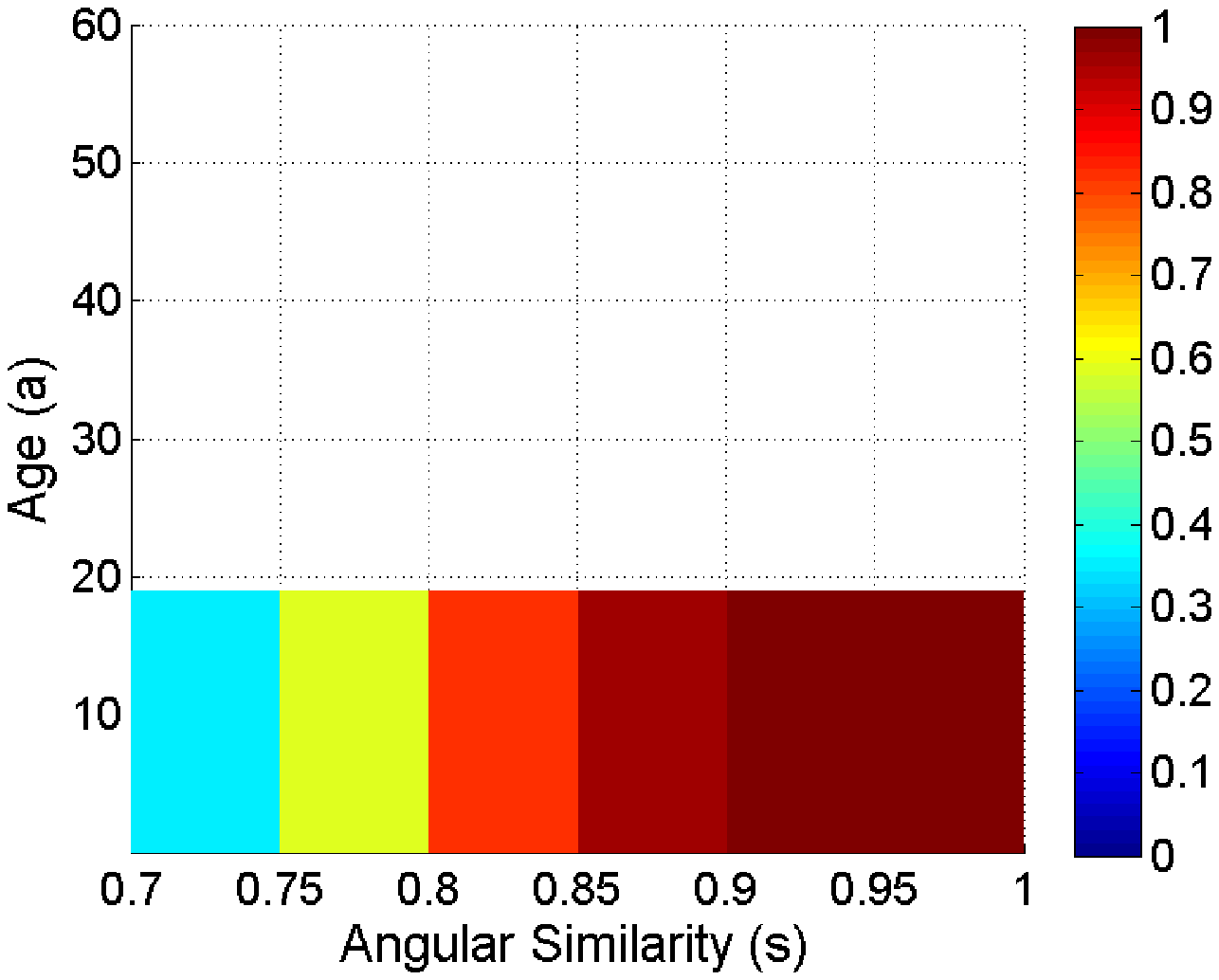}\label{fig:sp_agelsh}}
        \subfloat[SP Smooth(10,15)]{\includegraphics[scale=0.3]{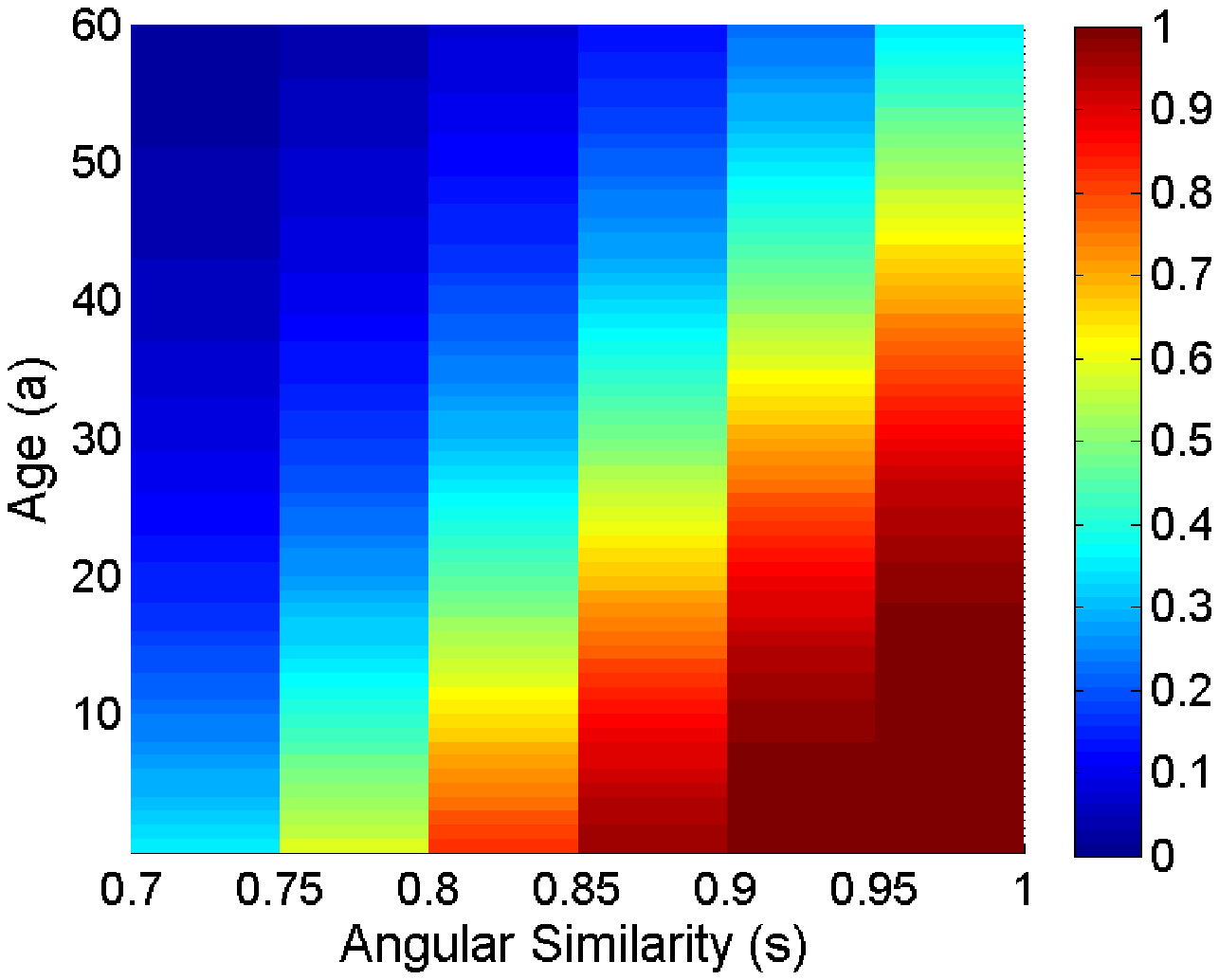}\label{fig:sp_mdlsh}}
        \vspace{1em}
        \caption{Success probability to find an item according to similarity and age
				         for a common index size.}
								\label{fig:spAnalytical}
\end{figure}

Figure \ref{fig:spAnalytical} illustrates as `heat maps' the success probabilities of Threshold and Smooth 
for this configuration.
The $x$ axis denotes similarity values $s$, and the $y$ axis denotes age values $a$.
Figure \ref{fig:spAnalytical}\subref{fig:sp_agelsh} depicts 
$SP(\mbox{Threshold}(10,15), s, a, 1)$,
while figure \ref{fig:spAnalytical}\subref{fig:sp_mdlsh} depicts
$SP(\mbox{Smooth}(10,15), s, a, 1)$.

As Threshold completely eliminates all item's copies that reach age $20$,
the success probability for $a \ge 20$ is $0$ (colored white).
The success probability of newer items behaves according to standard LSH,
i.e., the more similar an item is to the query ($s$ is closer to $1$),
the higher the success probability (color tends towards red).
Fixing an $s$ value, the success probability remains constant as $a$ increases,
since the number of buckets an item is indexed into remains constant.
With Smooth, on the other hand, for a fixed $s$ value, the success probability gradually decays as $a$ increases.
Smooth retains items for a longer time period than Threshold, 
and thus the success probability is non-zero for items older than $20$.

\paragraph*{Cumulative success probability}
We next formulate the \emph{cumulative success probability (CSP)} over similarity, age, and quality radii.
Given a query $q$, a similarity radius $R_{sim}$, an age radius $R_{age}$, and a quality radius $R_{quality}$,
CSP quantifies an algorithm's probability to find an item $x$ 
for which $\!sim(q,x) \in [R_{sim}, 1]$, 
$age(x) \in [0,R_{age}]$,
and $quality(x) \in [R_{quality}, 1]$.
CSP is the expected SP over all choices of 
$s \in [R_{sim}, 1]$, $a \in [0,R_{age}]$, and $z \in [0,R_{quality}]$
and is given by the following equation:

\begin{equation}
\label{def:rsp}
\begin{array}{l}
CSP(A, R_{sim}, R_{age}, R_{quality}) =  \nonumber \\
\ \ \ \ \displaystyle \int_{z=R_{quality}}^{1} \int_{a=0}^{R_{age}} \int_{s=R_{sim}}^{1} \frac{f(s,a,z) SP(A,s,a,z)}{\psi(R_{sim},R_{age},R_{quality})} dsdadz,
\end{array}
\end{equation}
where $f(s,a,z)$ denotes the joint probability density function of similarity $s$,
age $a$, and quality $z$, \\
and 
\begin{equation}
\begin{array}{l}
\psi(R_{sim},R_{age},R_{quality}) =  \nonumber \\
\ \ \ \ \displaystyle \int_{z=R_{quality}}^{1} \int_{a=0}^{R_{age}} \int_{s=R_{sim}}^{1} f(s,a,z) dsdadz,
\end{array}
\end{equation}
is a normalization factor.

Threshold keeps items up to age $\frac{1}{1-p}$;
from \eqref{eq:sp_agelsh} and \eqref{eq:sp_mdlsh}:
\begin{equation}
\begin{array}{l}
CSP(Threshold(k,L), R_{sim}, R_{age}, R_{quality}) = \nonumber \\
\ \ \displaystyle  \int_{z=R_{quality}}^{1} \int_{a=0}^{\frac{1}{1-p}} \int_{s=R_{sim}}^{1} \frac{f(s,a,z) (1-(1-s^kz)^L)}{\psi(R_{sim},R_{age},R_{quality})}  dsdadz
\end{array}
\end{equation}
and

\begin{equation}
\begin{array}{l}
CSP(Smooth(k,L), R_{sim}, R_{age}, R_{quality}) = \nonumber \\
\ \ \displaystyle \int_{z=R_{quality}}^{1} \int_{a=0}^{R_{age}} \int_{s=R_{sim}}^{1} \frac{f(s,a,z) (1-(1-p^as^kz)^L)}{\psi(R_{sim},R_{age},R_{quality})} dsdadz
\end{array}
\end{equation}

\paragraph*{Numerical illustration}
We compare\\ $CSP(A,R_{sim}, R_{age}, R_{quality})$ of Stream-LSH with Threshold and Smooth.
We pose the following assumptions:
we focus on the effect of the retention policy and hence we assume a constant quality function $\forall x$ $quality(x)=1$.
In general, the items' similarity distribution is data-dependent,
here, we assume a uniform distribution.
We consider a discrete age distribution because time is partitioned into discrete time ticks.
We assume a constant number of items arriving at each time unit, hence items' age is distributed uniformly.
Last, we assume that similarity and age are independent.
Under these assumptions we get:

\begin{equation}
\begin{array}{l}
CSP(Threshold(k,L), R_{sim}, R_{age}) = \nonumber \\
\ \ \ \ \displaystyle \sum_{a=0}^{\frac{1}{1-p}} \int_{s=R_{sim}}^{1} \frac{(1-(1-s^k)^L)}{R_{age}(1-R_{sim})} ds,
\end{array}
\end{equation}
and

\begin{equation}
\begin{array}{l}
CSP(Smooth(k,L), R_{sim}, R_{age}) =  \nonumber \\
\ \ \ \ \displaystyle  \sum_{a=0}^{R_{age}} \int_{s=R_{sim}}^{1} \frac{(1-(1-p^as^k)^L)}{R_{age}(1-R_{sim})} ds.
\end{array}
\end{equation}

\begin{figure}[tbh]
        \centering
        \subfloat[$R_{sim}=0.8$]{\includegraphics[scale=0.3]{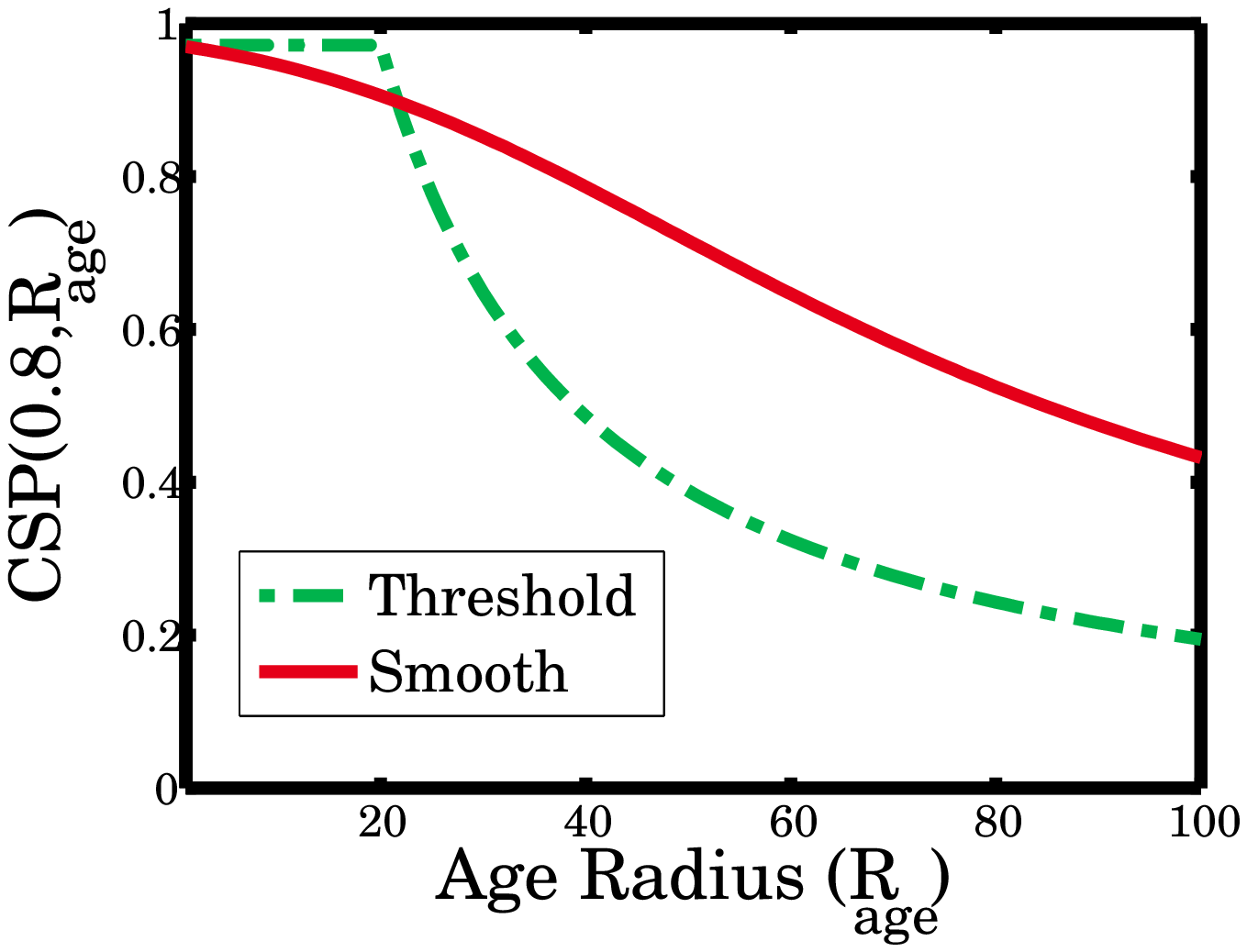}\label{fig:cspAnalyticalComaprison08}}
        \subfloat[$R_{sim}=0.9$]{\includegraphics[scale=0.3]{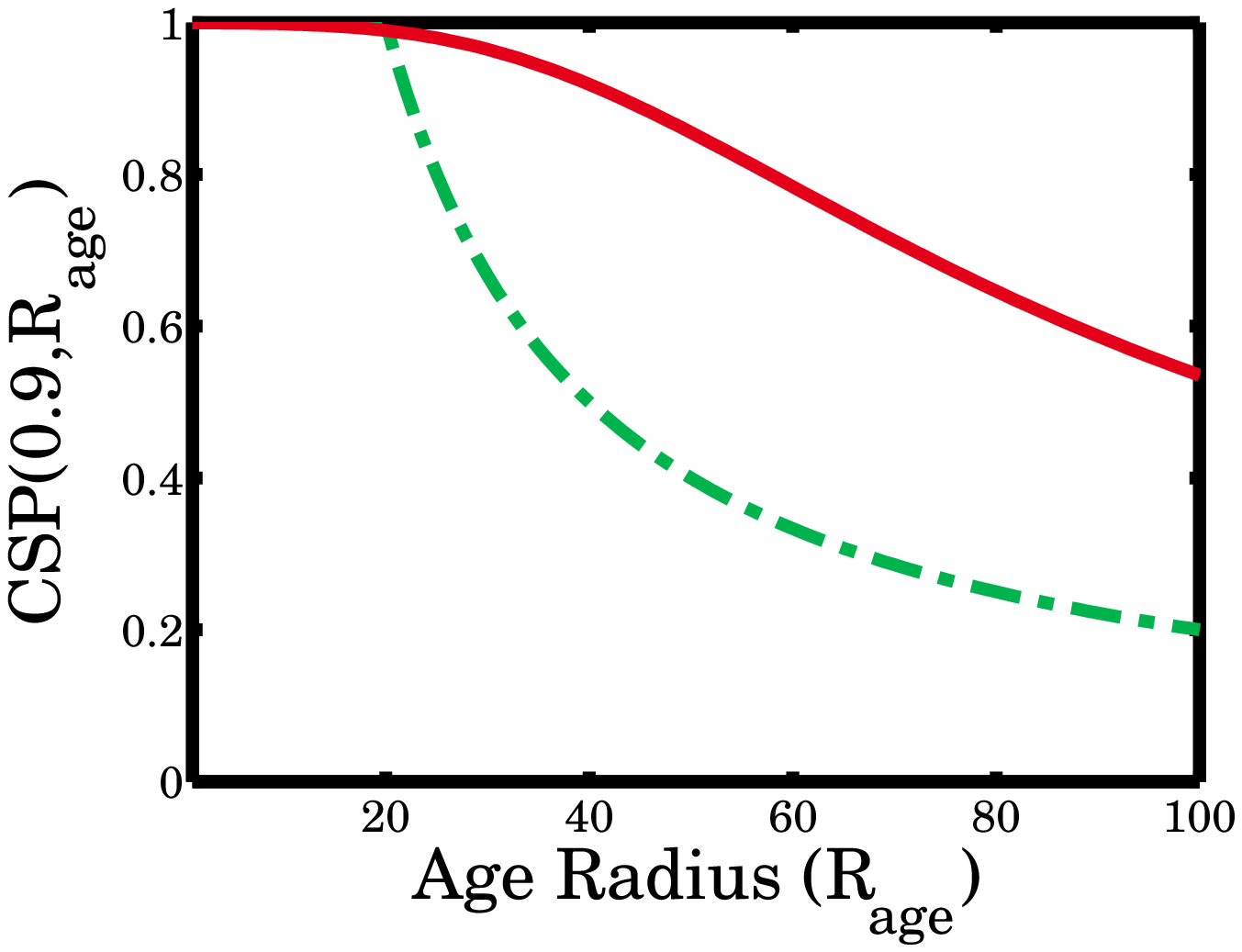}\label{fig:cspAnalyticalComaprison09}}   
        \vspace{1em}
\caption{Cumulative success probability of Threshold and Smooth by age radius for a common index size.}
		\label{fig:cspAnalyticalComaprison}
\end{figure}

For the purpose of the illustration, we select the same configuration as above:
$k=10$, $L=15$, $T_{size}=20\mu$, and $p=0.95$. 
Figures~\ref{fig:cspAnalyticalComaprison}\subref{fig:cspAnalyticalComaprison08} and~\ref{fig:cspAnalyticalComaprison}\subref{fig:cspAnalyticalComaprison09}
depict CSP for fixed $R_{sim}$ values $0.8$ and $0.9$, respectively,
and a varying age radius $R_{age}$.
The graphs show a freshness-similarity tradeoff between Threshold and Smooth.
Smooth has a better CSP for high age radii ($R_{age}$ exceeds $20$), for any $R_{sim}$ value.
This comes at the cost of a decreased CSP for similarity radius $0.8$
when $R_{age} \le 20$.

\COMMENT {
We proceed to examine how $L$ affects the results of Threshold and Smooth.
Figure \ref{fig:radiusSP} depicts their cumulative success probabilities for $R_{sim}=0.7$ and $R_{age}=60$,
as a function of $L$. 
For all $L$ values, the CSP of Smooth is greater than that of Threshold. 
The improvement increases as $L$ increases.
This implies that Smooth in some sense makes better use of additional resources it is provided with.

\begin{figure}[hbt]
    \centering
    \includegraphics[scale=0.3, clip]{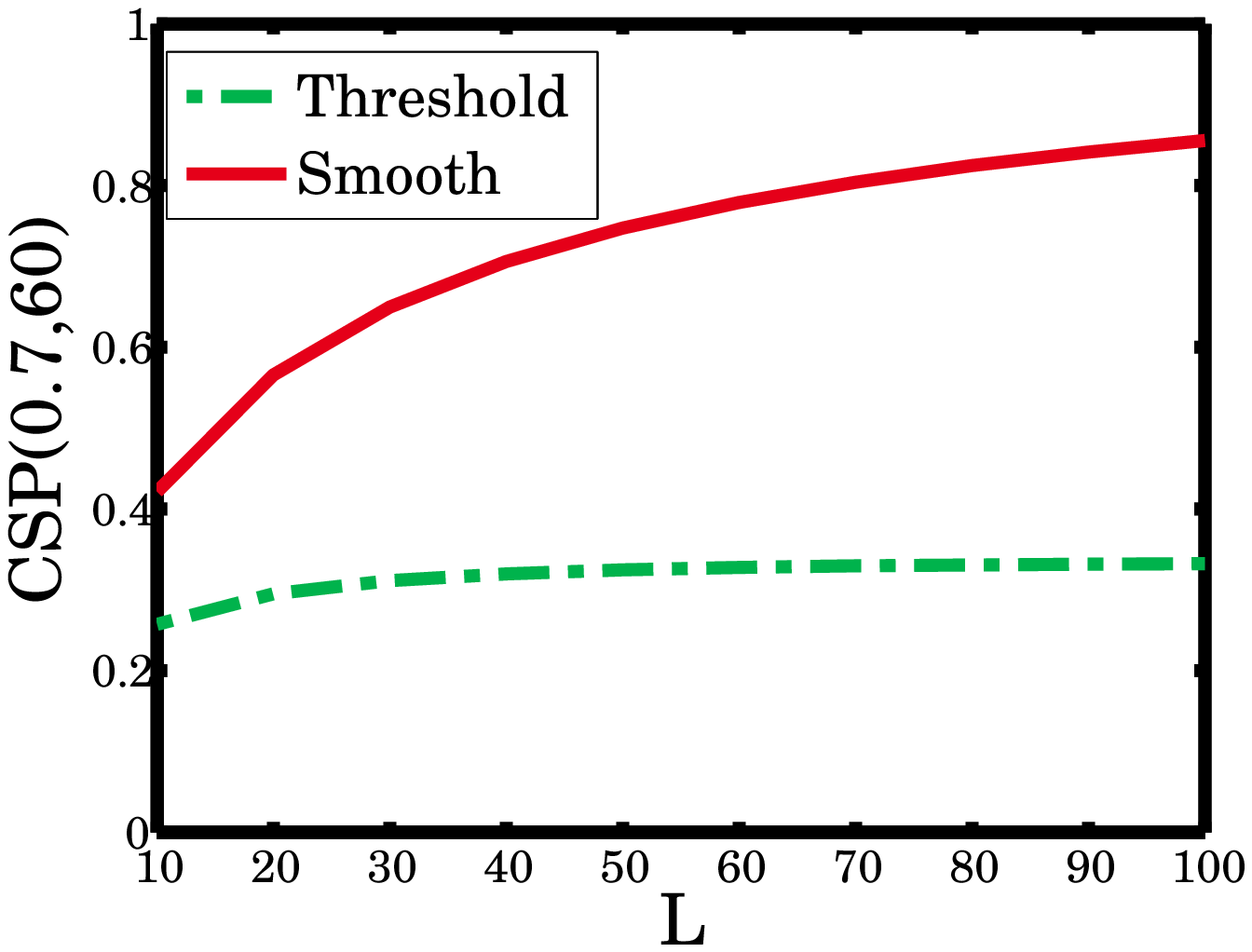}
    \vspace{1em}
    \caption{Analytical comparison of $\mbox{CSP(0.7, 60)}$ of Threshold and Smooth,
			for a fixed $k=10$, a varying $L$ parameter,
			and a common index size (Threshold and Smooth per $k,L$ pair). 
            }
			\label{fig:radiusSP}
\end{figure}
}

\subsubsection{Quality-Sensitivity}
Some applications, most notably social media ones, commonly handle content of varying quality,
and in particular, large amounts of low quality user-generated content~\cite{Agichtein2008,Becker11,EarlyBird}.
We show that for such applications, quality-sensitive indexing is expected to be attractive,
as it can better utilize space for improving the CSP of high quality items.
We compare Stream-LSH's quality-sensitive indexing,
which indexes an item with redundancy that is proportional to its quality,
to a quality-insensitive Stream-LSH,
which indexes $L$ copies of each item regardless of its quality.
We use the Smooth retention policy 
and the same index size for both variants.

\begin{figure}[hbt]
        \centering
        \subfloat[$R_{quality}=0.5$]{\includegraphics[scale=0.3]{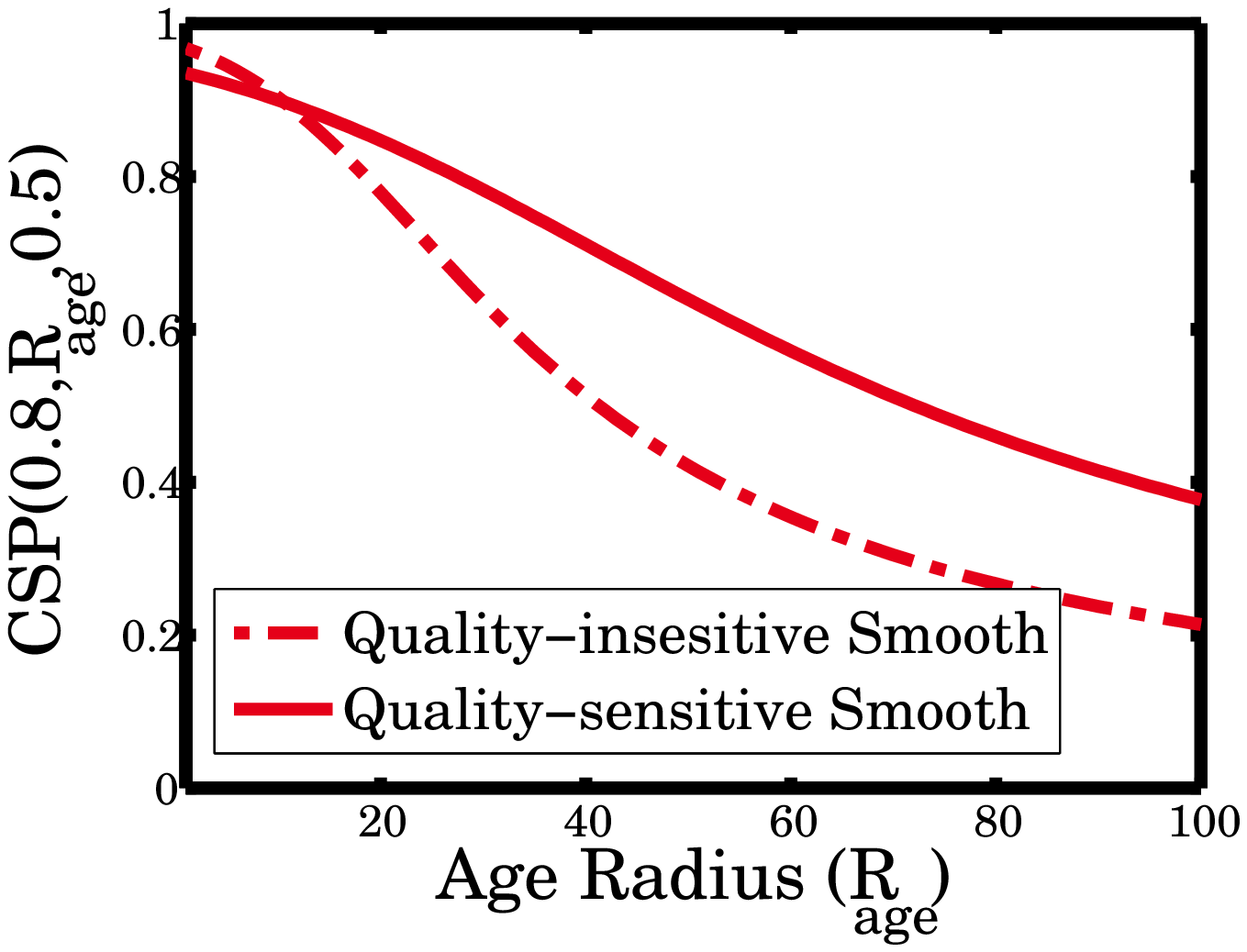}\label{fig:cspQuality_medium}}
        \subfloat[$R_{quality}=0.9$]{\includegraphics[scale=0.3]{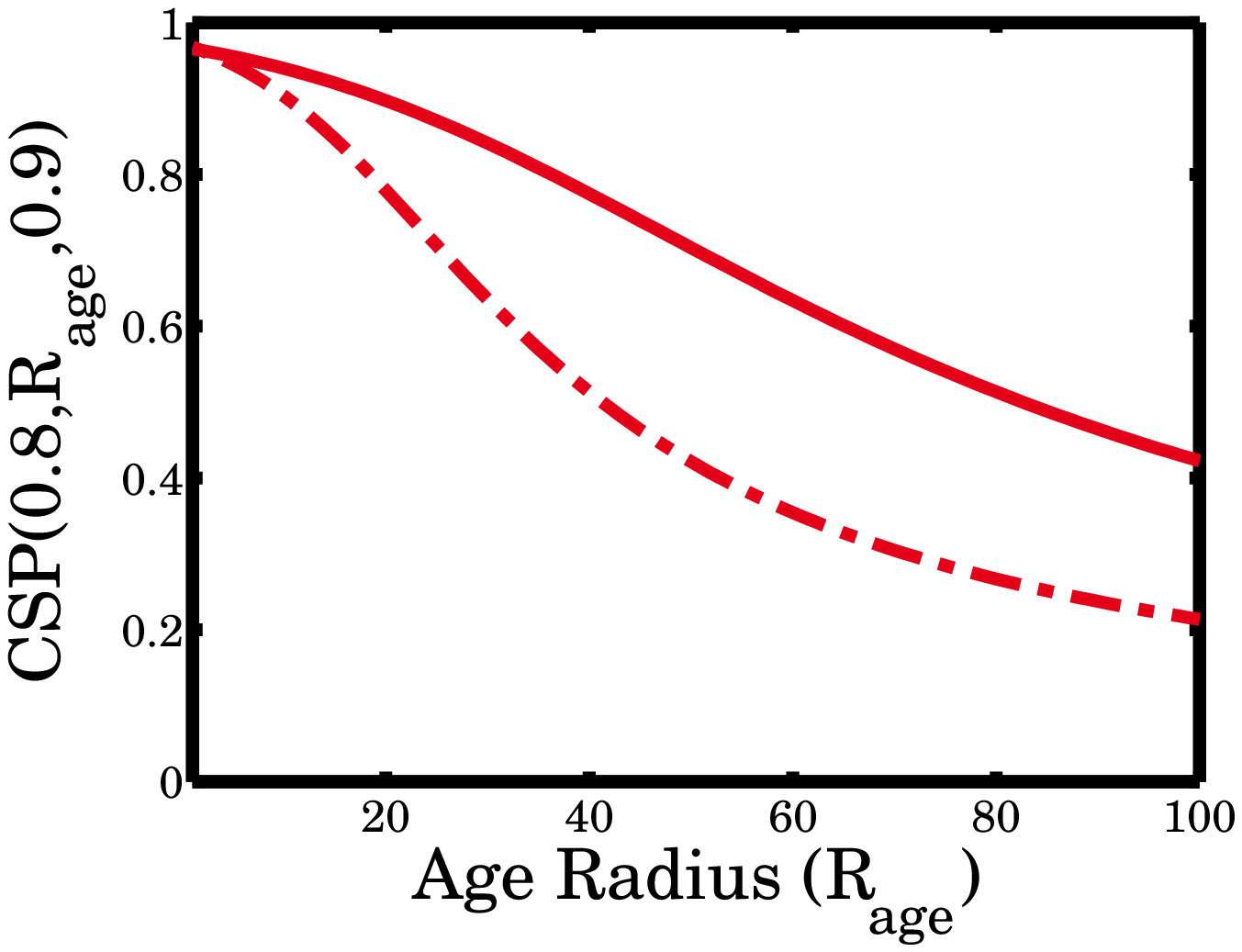}\label{fig:cspQuality_high}}  
        \vspace{1em}
        \caption{Cumulative success probability comparison of quality-sensitive and quality-insensitive Stream-LSH by age radius for a common index size.}
\label{fig:cspQuality}
\end{figure}

Assuming an average quality of $0.5$,
the expected number of newly indexed items per table 
is $\mu$ according to quality-insensitive indexing,
and $0.5\mu$ according to quality-sensitive indexing.
To obtain a common index size of $10\mu L$,
we use $p=0.9$ in the quality-insensitive algorithm
and $p=0.95$ in the quality-sensitive one (Proposition \ref{prop:plshTableSize}).
Figure~\ref{fig:cspQuality} illustrates the two Stream-LSH variants for $R_{sim}=0.8$
and varying $R_{age}$ radii.
In Figure~\ref{fig:cspQuality}\subref{fig:cspQuality_medium}, we examine items above the mean quality ($R_{quality}=0.5$),
and in Figure~\ref{fig:cspQuality}\subref{fig:cspQuality_high} we examine high-quality items ($R_{quality}=0.9$).
Both figures show that quality-sensitive indexing
increases the CSP compared to the quality-insensitive variant.
The improvement is even more marked for high-quality items ($R_{quality}=0.9$).

\subsubsection{Dynamic Popularity}
\label{sec:analysisDynPop}
We next analyze an item's success probability according to 
Stream-LSH when using \DPOP{} and the Smooth retention policy. 
We first formulate the \emph{bucket probability (SB)} to find an item in a bucket it is hashed to.
We then use SB to formulate an item's success probability.

For the sake of the analysis, 
we assume that the interest in an item 
does not vary over time.
At each time $t_i$, $x$ is included in the interest stream with some 
probability $\rho_x$, which we call the item's \emph{interest probability}. 
That is, for all $t_i$, x appears with probability
$\rho_x$.
According to Definition~\ref{def:popScore} and due to the linearity of expectation:
\[
E(pop(x))=(1-\alpha) \sum_{i=0}^n E(a_i(x)) \alpha^{(n-i)},
\]
which is a geometrical series that converges to $E(a_i(x))$ when $n \rightarrow \infty$.
As our stream is infinite and $E(a_i(x))=\rho_x$:
\begin{equation}
\label{eq:exp_pop}
E(pop(x))=\rho_x.
\end{equation}
When clear from the context, we omit $x$ and denote $\rho$ for brevity.

\paragraph*{Bucket probability}
We denote by $SB(p, u, \rho, z, n)$ the probability that $x$ is stored in its bucket at time $t_n$,
where $u$ and $p$ are the insertion and retention factors, respectively,
$quality(x)=z$,
and $\rho$ is $x$'s interest probability.
We denote by $E_i$, $0 \le i \le n$, the event that $x$ is inserted to its bucket at time $t_i$,
and survives elimination until time $t_n$, but is not selected for insertion to its bucket at any subsequent time 
$t_j$, $i < j \le n$. Then
\[
Pr(E_i) = zu\rho p^{(n-i)} (1-zu\rho)^{n-i} = zu\rho [p(1-zu\rho)]^{n-i}
\]
Since $SB(p, u, \rho, z, n) = Pr(\bigcup_{i=0}^n E_i)$
and $\bigcup_{i=0}^n E_i$ is a union of pairwise disjoint events, 
it follows that 
\[
SB(p, u, \rho, z, n) = \sum_{i=0}^n zu\rho [p(1-zu\rho)]^{n-i}.
\]
$SB(p, u, \rho, z, n)$ is a geometric series that converges to 
$\frac{zu\rho}{1-p(1-zu\rho)}$ when $n \rightarrow \infty$.
Our interest stream is infinite, thus:
\begin{prop}
\label{prop:sb_dynapop}
Given an item $x$, 
the probability $SB(p, u, \rho, z)$ to find item $x$ in its bucket when
using Stream-LSH with \DPOP{} and the Smooth retention policy is $\frac{zu\rho}{1-p(1-zu\rho)}$,
where $u$ and $p$ are the algorithm's insertion and retention factors respectively, 
$quality(x)=z$,
and $\rho$ is $x$'s interest probability.
\end{prop}

\paragraph*{Numerical illustration}
We illustrate SB for an interest probability that follows a Zipf distribution
(typical in social phenomena~\cite{TI2011})
which implies a small number of very popular items
and a long tail of rare ones.
We consider a Zipf distribution where
the $r$-th ranked item $x$ has an interest probability $\rho_x=1/r$.
We set the quality $z$ to $1$.

\begin{figure}[hbt]
    \centering
		\subfloat[Insertion factor $u$]{\includegraphics[scale=0.3, clip]{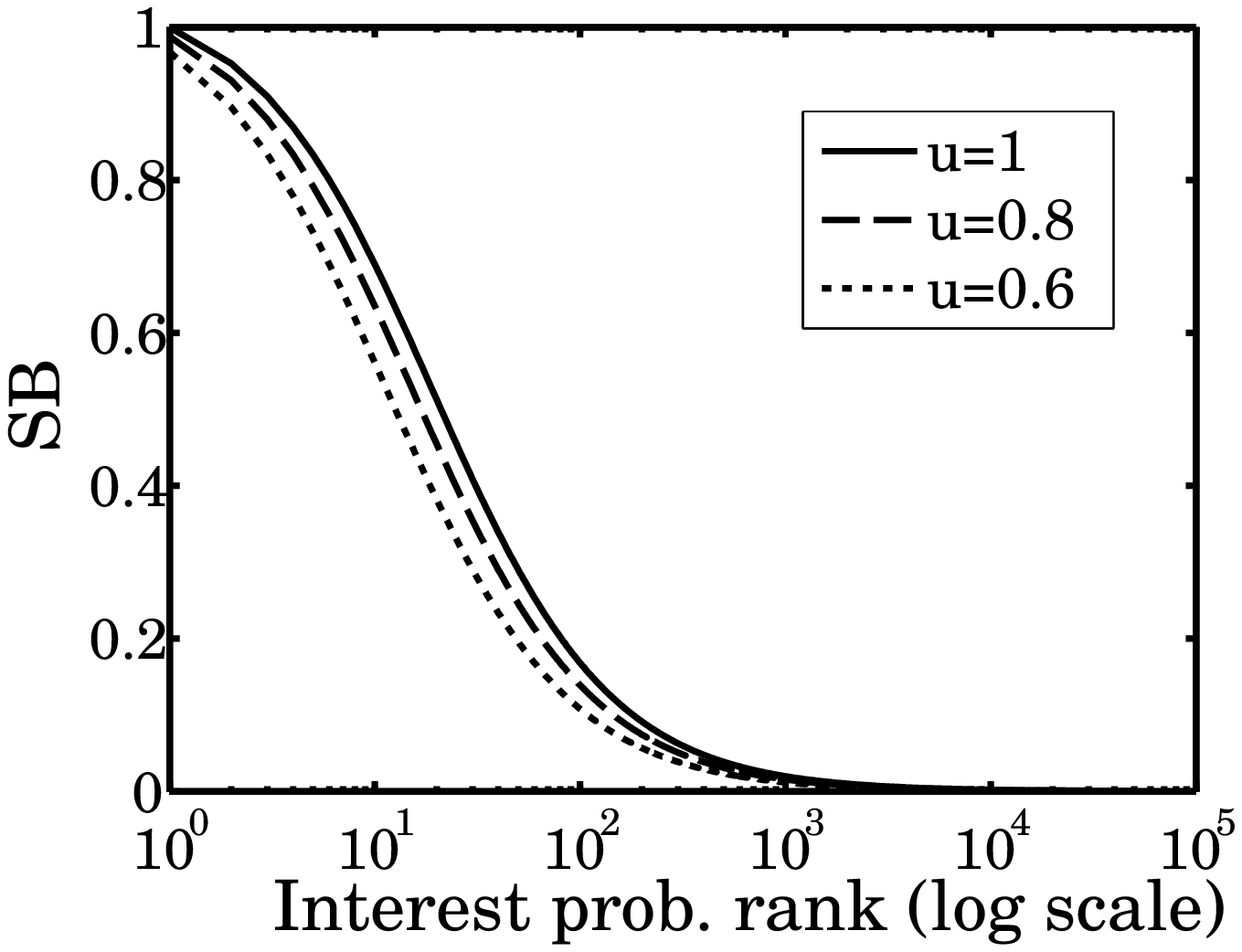}\label{subfig:sb_u}}
		\subfloat[Retention factor $p$]{\includegraphics[scale=0.3, clip]{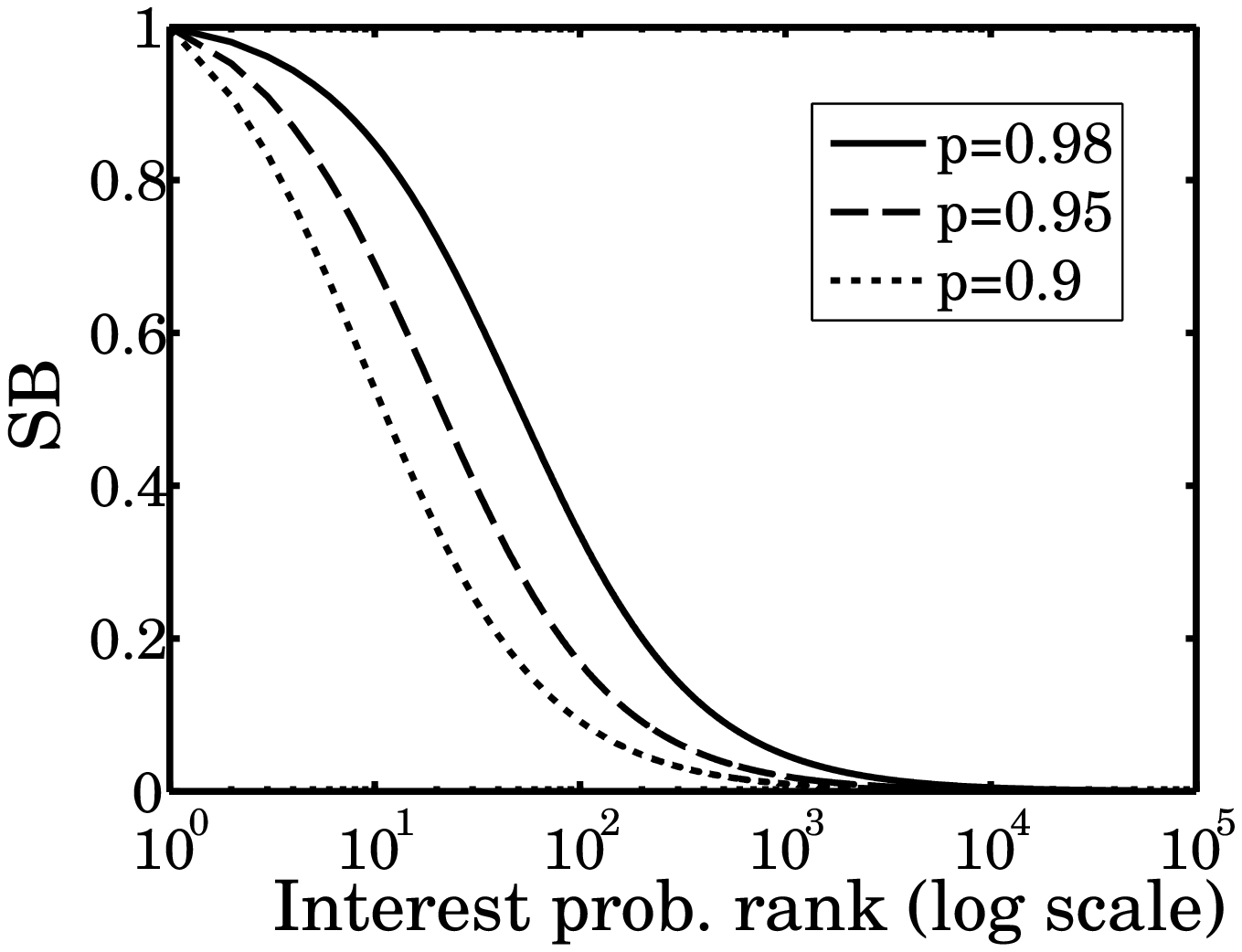}\label{subfig:sb_p}}
        \vspace{1em}
    \caption{Effect of parameters on the probability to find an item in its bucket according to \DPOP{}.}    
		\label{fig:sb_params}
\end{figure}

Figure \ref{fig:sb_params} illustrates the SB according to interest probability rank, 
for different values of $u$ and $p$.
In Figure \ref{fig:sb_params}\subref{subfig:sb_u} we fix $p=0.95$ and examine the effect of the insertion probability $u$.
The graphs illustrate that increasing $u$ increases SB most notably for popular items.
In Figure \ref{fig:sb_params}\subref{subfig:sb_p} we fix $u=1$ and examine the effect of the retention probability $p$.
The graphs illustrate that when $p$ increases, \DPOP{} retains additional items of lower popularity.

\paragraph*{Success probability}
\label{sec:spDynPop}
We denote by $SP(\DPOP{}(k,L), s, w, z)$ 
the probability of Stream-LSH with \DPOP{} and the Smooth retention policy to find an item $x$ for query $q$,
s.t. $sim(q,x)=s$, $E(pop(x))=w$, and $quality(x) = z$.
By applying Proposition \ref{prop:sb_dynapop} and as $w=\rho$ (Equation \ref{eq:exp_pop}):
\begin{equation}
SP(\DPOP{}(k,L), s, w, z)=1-(1-\frac{zuw}{1-p(1-zuw)}s^k)^L
\end{equation}
The cumulative success probability is computed similarly to the cumulative success probability analysis 
in Section \ref{sec:sp_rPolicies},
and in particular depends on the distribution of $w$. 

\begin{figure}[hbt]
    \centering
    \includegraphics[scale=0.4, clip]{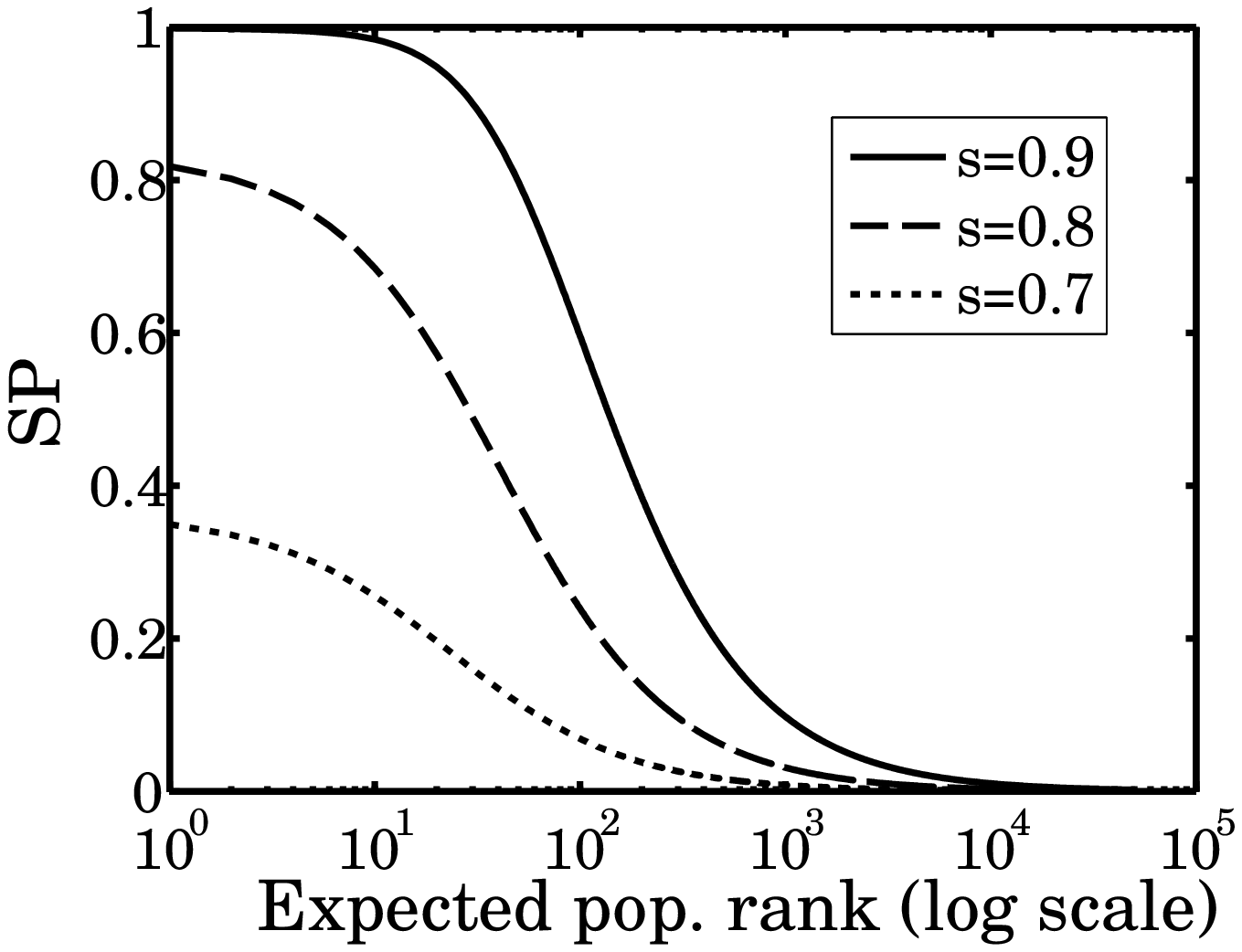}
    \vspace{1em}
    \caption{Success probability to find an item as a function of its expected popularity according to \DPOP{}.}    
						\label{fig:sp_analysis1}
\end{figure}

\paragraph*{Numerical illustration}
Figure \ref{fig:sp_analysis1} depicts $SP(\DPOP{}(k,L), s, w, z)$ as a function of $w$'s rank.
We illustrate SP for three $s$ values: $0.7$, $0.8$, and $0.9$.
We fix $k=10$, $L=15$, $z=1$, $p=0.95$ and $u=1$.
As the graphs show, 
an item's success probability increases as its similarity to the query increases.
Additionally, an item's success probability increases as its expected popularity increases.

\section{Empirical Study}
\label{sec:eval}
We conduct an empirical study of Stream-LSH using real world stream datasets
and evaluate its effectiveness using the recall metric.

\subsection{Methodology}
\label{sec:methodology}
\paragraph*{External libraries}
We use Apache Lucene 4.3.0~\cite{Lucene} search library for the indexing and retrieval infrastructure.
For retrieval, we override Lucene's default similarity function by implementing angular 
similarity according to Equation \ref{eq:ang}. 
For the LSH family of functions, we use TarsosLSH~\cite{TarsosLSH}.

\paragraph*{Datasets}
We use Reuters RCV1~\cite{Reuters} news dataset and Twitter~\cite{TweeterSNAP,SSSJ16} social dataset.
In both datasets, each item is associated with a timestamp denoting its arrival time.
The Reuters dataset consists of news items from August 1996 to August 1997,
and the Twitter dataset consists of Tweets collected in June 2009.
These datasets do not contain quality information and so we assume 
$quality(x)=1$ for all items.
In order to evaluate quality-sensitivity,
we use a smaller Twitter dataset~\cite{TwitterNas}, denoted TwitterNas, 
consisting of a stream of Nasdaq related Tweets spanning $97$ days from March 10th to June 15th 2016.
TwitterNas contains number of followers of Tweets authors, which we use for 
assigning quality scores to Tweets (see Section \ref{sec:eQuality}).
In all datasets, 
we represent an item as a (sparse) vector whose dimension is the number of unique terms in the entire dataset,
and each vector entry corresponds to a unique term,
weighted according to Lucene's $\mbox{TF-IDF}$ formula.
\COMMENT {
where $TF(term)$ is the square root of the term's frequency in the item, 
and $IDF(Term) = ln(\frac{N_d}{N_{Term}+1})+1$, 
where $N_d$ is the total number of items,
and $N_{Term}$ is the number of items containing the term.
While there exist a variety of methods for representing textual items~\cite{IRBook},
the choice of method is orthogonal to our research,
and so we simply stick to Lucene's conventional approach.
Table \ref{table:stats} summarizes the datasets' statistics.
}

\paragraph*{Train and test}
We partition each dataset into (disjoint) train and test sets.
The train set is the prefix of the item stream up to a tick that we consider to be the current time.
The test set is the remainder of the dataset,
which was not previously seen by the Stream-LSH algorithm.
We randomly sample an evaluation set $Q$ of $3,\!000$ items from the test set 
and compute recall over $Q$ according to the given radii. 
Table \ref{table:stats} summarizes the train and test statistics.

\begin{table*}[hbt]
\begin{center}
\begin{tabular}{|c|c|c|c|c|c|}
\hline
       		  &              &  \multicolumn{2}{|c|}{Train}       & \multicolumn{2}{|c|}{Test}     \\ \hline 
              & Time unit    &  Num. items         & Num. ticks   & Num. items  & Num. ticks  \\ \hline 
Reuters       & Day          &  $756,\!927$        & $343$            & 22,\!986     &  10 \\                   
Twitter       & 10  Minutes  &  $18,\!224,\!293$   & $2,\!705$       & 42,\!296      & 10 \\                   
TwitterNas    & Day          &  $275,\!946$         & $92$            & 18,\!831      & 5 \\ \hline

\end{tabular}
\vspace{1em}
\caption{Train and test statistics.}
\label{table:stats}
\end{center}
\end{table*}

\COMMENT {
\paragraph*{Simulation}
Given $k$ and $L$, we use TarsosLSH for randomly selecting $L$ hash functions 
with domain $\left\{0,1\right\}^k$.
For each of the retention policies,
we implement an LSH index that consists of $L$ hash tables, 
and use the hash functions for mapping vectors to their buckets. 
We maintain the tables in a Lucene index so we can efficiently search over them at runtime.
We simulate Stream-LSH by running Algorithm \ref{alg:stream-lsh} over the training set in order of age.
For each dataset, we define a time unit that yields an arrival rate of several thousands of items per tick.
For the exact search, we construct a standard Lucene index of the entire training set.

Given a query $q \in Q$, we retrieve its ideal result set,\\ Ideal$(q,R_{sim},R_{age})$, 
by searching for all items within the requested similarity and age radii in the standard index.
For each retention policy, we search the query's buckets
for items within the requested similarity and age radii,
and build $q$'s approximate result set, Appx$(A,q,R_{sim},R_{age})$,
as the union of the items found in these buckets.
For the quality-sensitive evaluation, we also consider a quality radius.
}

\subsection{Retention Policies}
\label{sec:eCSP}
We evaluate the recall of the three retention policies for the Reuters and Twitter datasets as a function of age.
As the retention aspect of our algorithm is orthogonal to the quality-sensitive indexing aspect, 
we assume here that $quality(x)=1$ for all items.
\COMMENT {
$Recall(A,R_{sim},R_{age})$ measures the fraction of the ideal result set 
that algorithm $A$ retrieves within $R_{sim}$ and $R_{age}$,
averaged over query set $Q$ (see Definition \ref{def:algRecall}).
Whereas our theoretical estimation of $CSP(A, R_{sim}, R_{age})$ 
in Section \ref{sec:csp}
assumes uniform distributions of similarity and age,
the empirical results depend on the actual data distribution,
which is not necessarily uniform.
}
In order to achieve a fair comparison, we use $k=10$ and $L=15$ for the three retention policies,
and configure them to use approximately the same index size: 
We set $T_{size}=45,\!000$ and $B_{size}=45$ in Reuters;
$T_{size}=180,\!000$ and $B_{size}=177$ in Twitter;
$p=0.95$ in both datasets.
\COMMENT {
Table \ref{table:config} summarizes our evaluation configuration.
\begin{table}[hbt]
\begin{center}
\begin{tabular}{|c|c|c|c|c|c|c|c|c}
\hline
       		  & K     & L  & Retention    & Bucket         & Appx. Avg.      \\ 
              &       &    & Factor       & Size Limit     & Table Size      \\ \hline 
Reuters       & 10    & 15 & $0.95$       & $45$           & $45,\!000$      \\ 
Twitter       & 10    & 15 & $0.95$       & $177$          & $180,\!000$     \\ \hline

\end{tabular}
\vspace{1em}
\caption{Retention policies evaluation configuration.}
\label{table:config}
\end{center}
\end{table}
}

\begin{figure}[hbt]
	\centering
    \subfloat[Reuters: $R_{sim}=0.8$]{\includegraphics[scale=0.3]{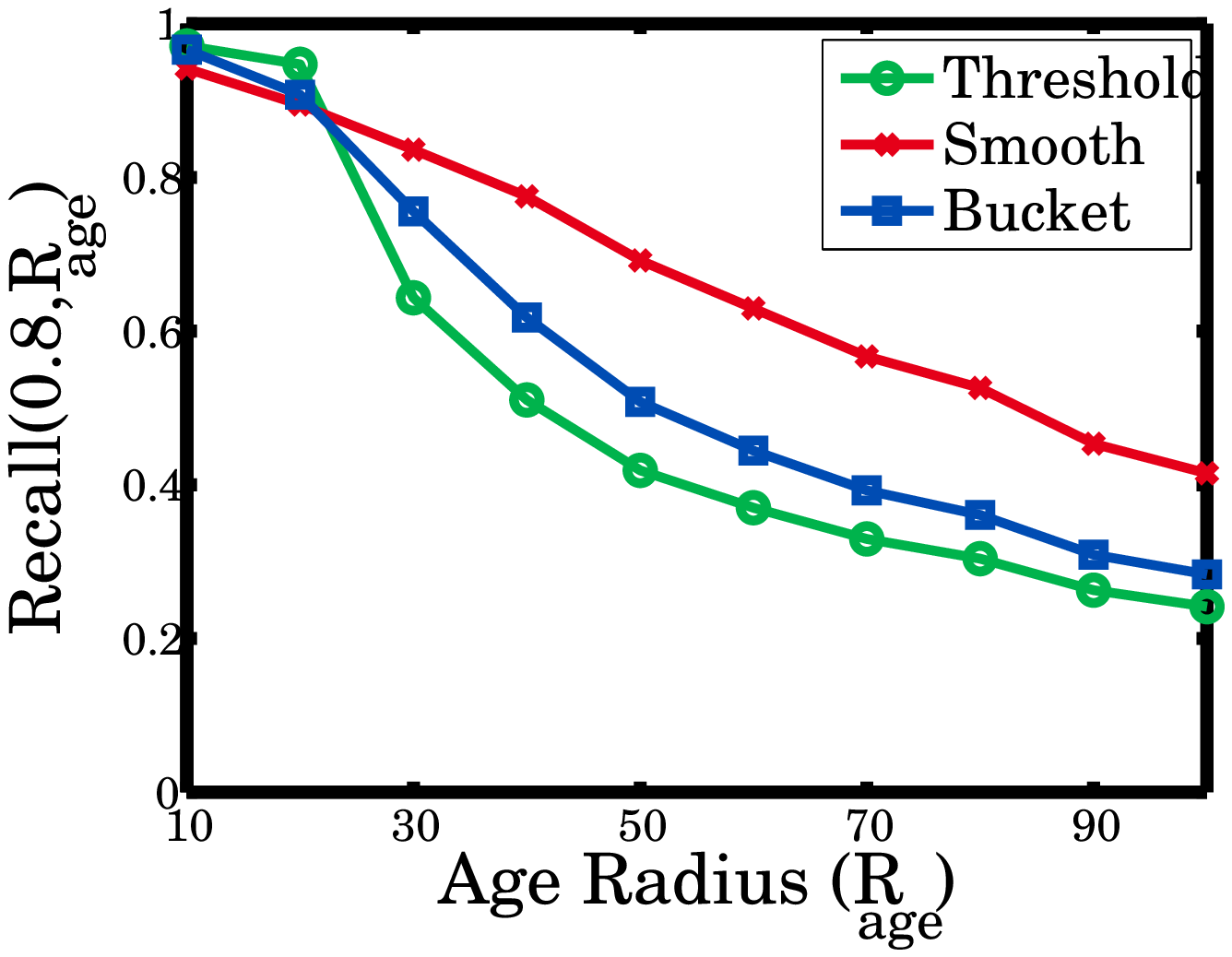}}
	\subfloat[Reuters: $R_{sim}=0.9$]{\includegraphics[scale=0.3]{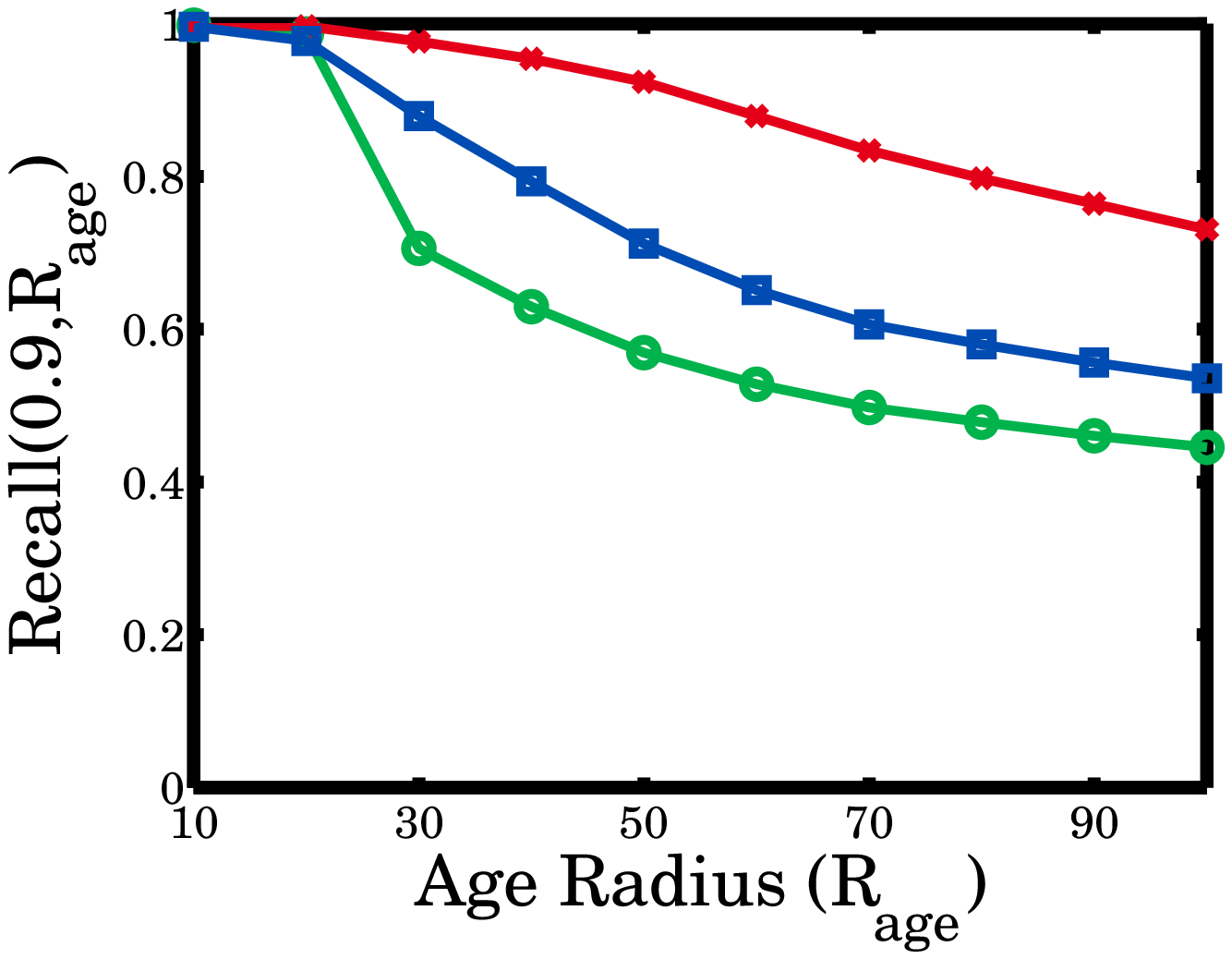}}
					   
    \subfloat[Twitter: $R_{sim}=0.8$]{\includegraphics[scale=0.3]{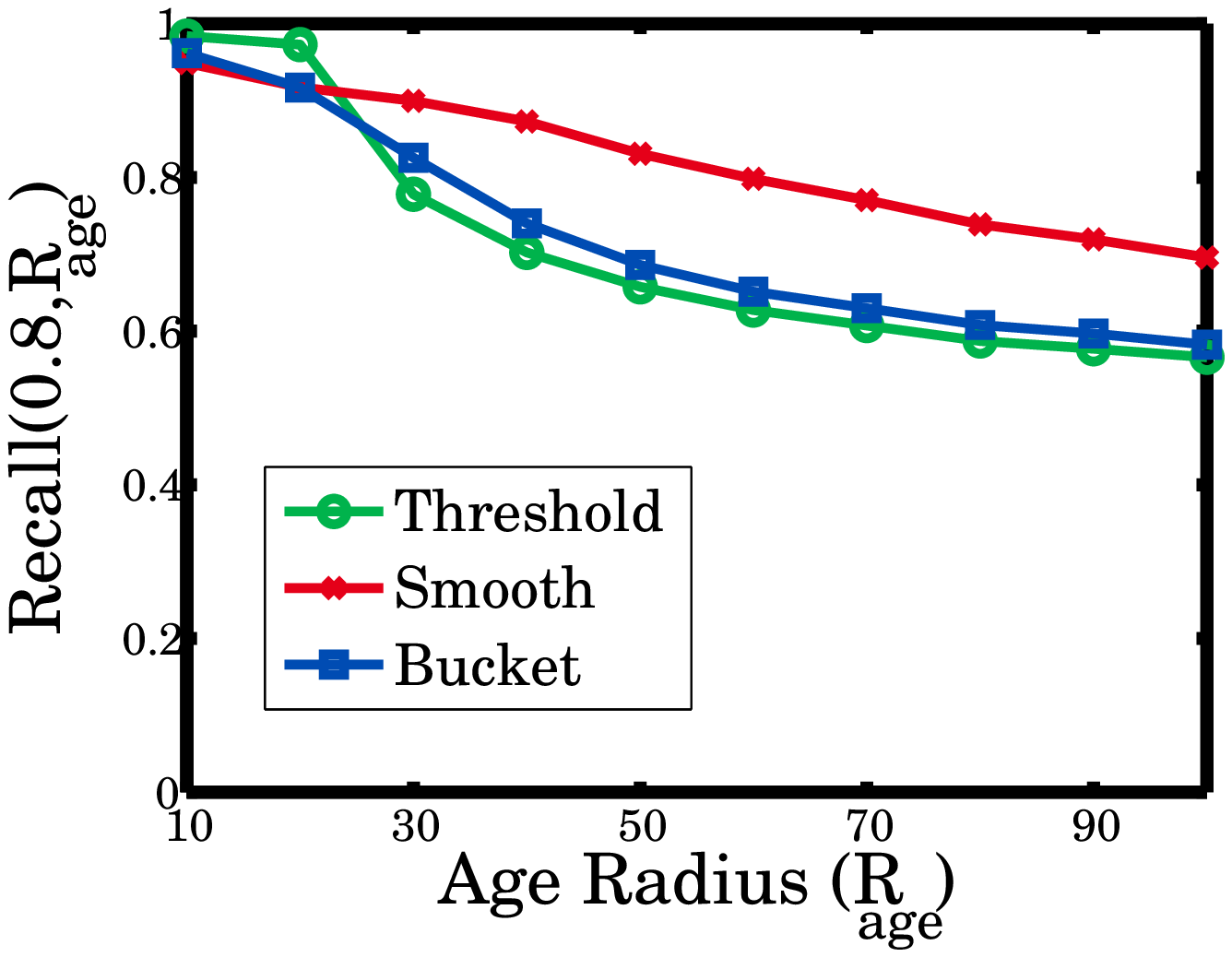}}
		\subfloat[Twitter: $R_{sim}=0.9$]{\includegraphics[scale=0.3]{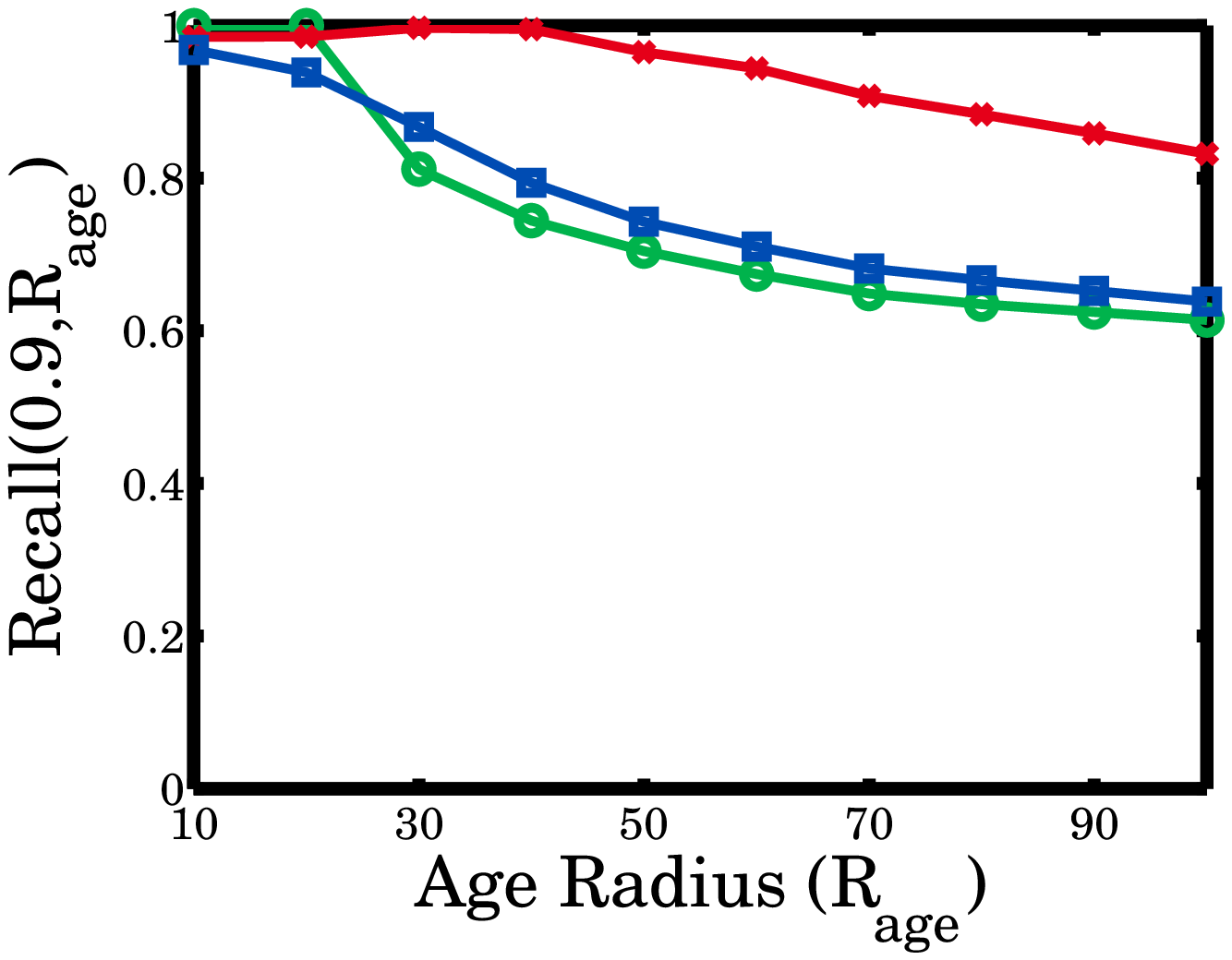}}
    \vspace{1em}
    \caption{Recall comparison by age radius of the three retention policies using approximately the same index size.}
								\label{fig:cspEmpComaprison}
\end{figure}

Figure \ref{fig:cspEmpComaprison} depicts our recall results
for Reuters in the top row, and Twitter in the bottom row.
Our goal is to retrieve items that are similar to the query,
hence we focus on $R_{sim}$ values $0.8$, and $0.9$.
As we are also interested in the retrieval of items that are not highly fresh,
we evaluate recall over varying age radii values.

When considering $R_{sim}=0.8$ (leftmost column) there is a tradeoff between Threshold and Smooth:
when focusing on the highly fresh items ($R_{age} < 20$),
Threshold's recall is slightly larger than Smooth's.
Indeed, Threshold is effective when only the retrieval of the highly fresh items is desired.
However, Smooth outperforms Threshold when the age radius increases to include also less fresh items.
For example, in Reuters, Smooth achieves a recall of $0.69$ for items that are at least $0.8$-similar to the query
and are not older than age $50$, and Threshold achieves a lower recall of $0.42$.
Bucket's recall is higher than Threshold's for ages that exceed $20$,
as unlike Threshold, Bucket does not eliminate items at once.
Yet, Smooth outperforms Bucket when increasing the age radius due to 
applying an explicit gradual elimination over all items.
When increasing the similarity radius to $R_{sim}=0.9$ (leftmost column),
the advantage of Smooth over Threshold becomes pronounced.
For example, in Twitter, Smooth achieves a recall of $0.97$ for items 
that are not older than age $50$, 
whereas Threshold only achieves a recall of $0.7$.

\subsection{Quality-Sensitivity}
\label{sec:eQuality}
We move on to evaluating Stream-LSH's quality-sensitive approach.
We experiment with the TwitterNas dataset,
which contains for each Tweet $x$
the number of followers of its author representing its authority, and denoted $T_f(x)$.
We define the following quality scoring function:
\[
quality(x) = log_2(1+min(1,T_f(x)/N_f)),
\]
where $N_f$ is a configurable normalization factor.
In our experiments, we set $N_f=5,\!000$ ($15\%$ of the authors have more than $5,\!000$ followers).
Applying $quality(x)$ on TwitterNas entails an average quality score of $0.33$.

\begin{figure}[hbt]
        \centering
        \subfloat[$R_{quality}=0.5$]{\includegraphics[scale=0.3]{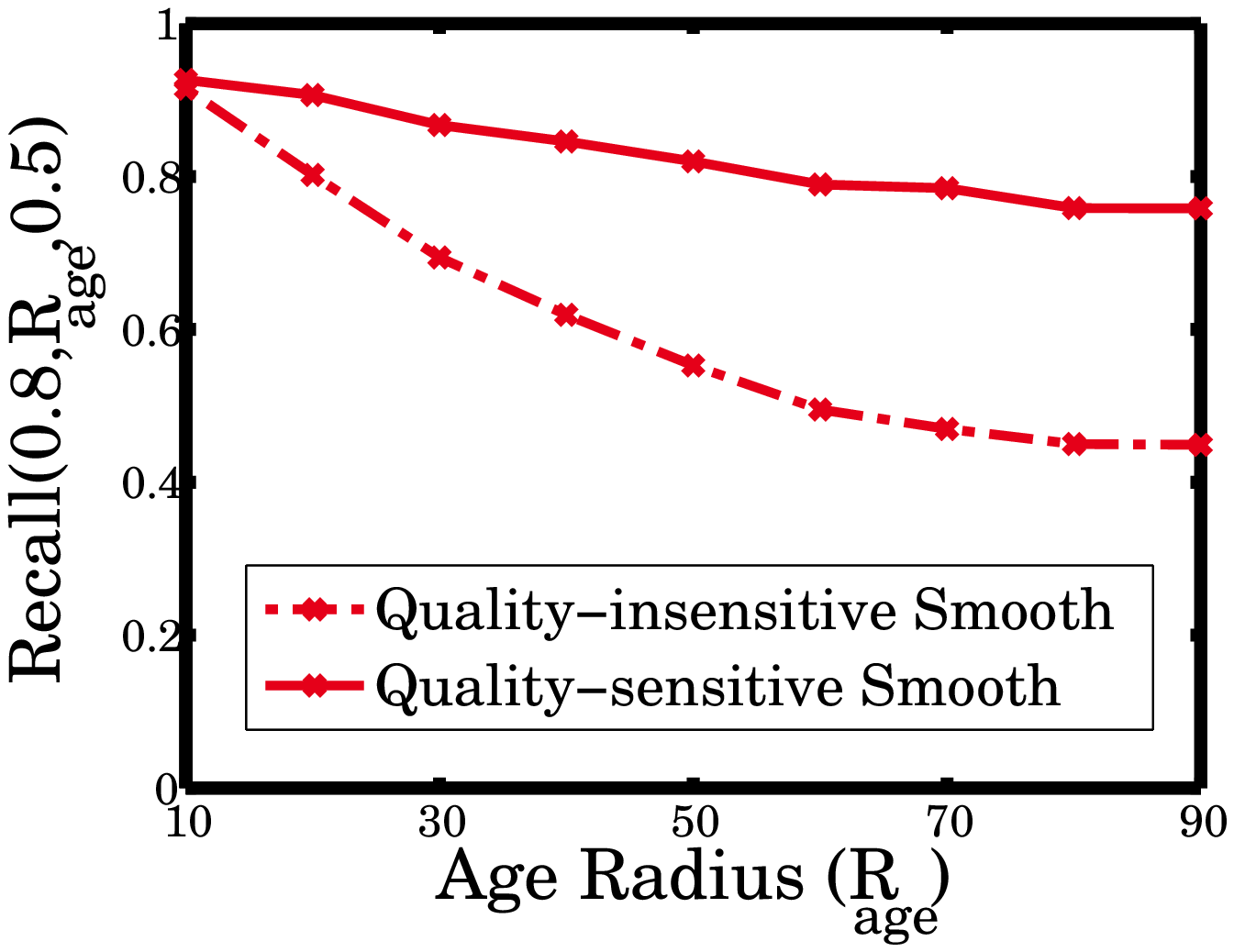}\label{fig:recallQuality_medium}}
        \subfloat[$R_{quality}=0.9$]{\includegraphics[scale=0.3]{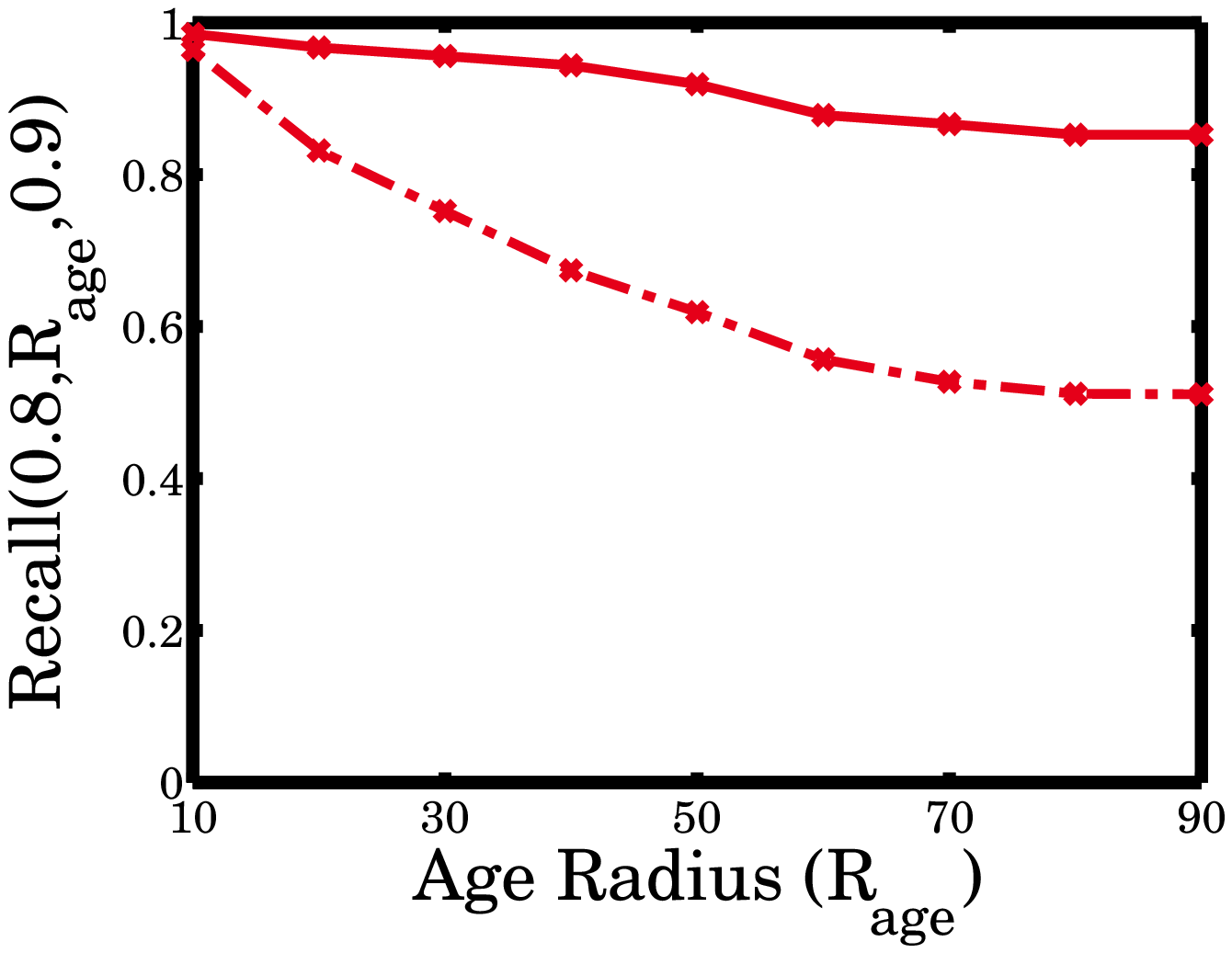}\label{fig:recallQuality_high}}          
\vspace{1em}
        \caption{Recall comparison of quality-insensitive and quality-sensitive Smooth using approximately the same index size.}
\label{fig:recallQuality}
\end{figure}

We experiment with quality-sensitive and quality-insensitive variants of Smooth,
with $k=10$ and $L=15$.
In order to conduct a fair comparison, 
we set retention factors that entail approximately the same index size for both variants.
More specifically, we set $p=0.9$ for the quality-insensitive variant, 
which results in an index size of $636,\!290$ items in our experiment,
and $p=0.97$ for the quality-sensitive variant which results in 
an index size of $590,\!818$ items in our experiment.
Recall that quality-sensitive indexing is more compact,
which enables a slower removal of item copies. 
This is reflected by a larger $p$ value in the quality-insensitive case.
We experiment with the same radii as in our analysis:
we fix $R_{sim} = 0.8$, and experiment with $R_{quality}=0.5$, and $R_{quality}=0.9$ over varying age values.
Figure \ref{fig:recallQuality} depicts the recall achieved by the two Smooth variants as a function of the age radius.

The graphs demonstrate that for both $R_{quality}$ values, the quality-sensitive approach significantly outperforms the quality-insensitive approach when searching for similar items ($R_{sim}=0.8$)
over all age radii values that we examined.
This is since the quality-sensitive approach better exploits the space resources for high quality items.
The advantage of quality-sensitive indexing increases as the age of high-quality items increases,
which is an advantage when the retrieval of items that are not necessarily the most fresh ones is desired.
For example, 
considering $R_{quality}=0.5$ (Figure \ref{fig:recallQuality}\subref{fig:recallQuality_medium}) and $R_{age}=30$, 
the recall achieved by quality-insensitive Smooth is $0.7$, 
whereas the recall achieved by quality-sensitive Smooth is $0.88$. 
When considering $R_{age}=90$, 
the recall of quality-insensitive Smooth is $0.45$,
whereas the recall of quality-sensitive Smooth is $0.76$.
A similar trend is observed for $R_{quality}=0.9$ (Figure \ref{fig:recallQuality}\subref{fig:recallQuality_high}).
Note that the graphs of $R_{quality}=0.9$ and $R_{quality}=0.5$ differ 
due to the different distributions of similarity, age, and quality, 
when considering different radii values in real data.
As we noted in our analysis, the advantage of the quality-sensitive approaches is most 
pronounced when there exists a large amount of low quality items in the dataset.
Indeed, in our setting, $73\%$ of the items are assigned a quality value below $0.5$.
In such cases, using quality-sensitive Stream-LSH is expected to be appealing for similarity-search stream applications.

\subsection{Dynamic Popularity}
\label{sec:evalDynPop}
We wrap up by studying Stream-LSH when using \DPOP{} and the Smooth retention policy.
We experiment with $u=0.95$, $p=0.95$.
As our datasets do not contain temporal interest information,
we simulate an interest stream $I$ by considering query results as signals of interests in items~\cite{TI2011}, as follows:
We use the first $75\%$ items in the train set as the item stream $U$.
We construct a query set $Q^{\ast}$ by randomly sampling each item from the remaining $25\%$ of the train set
with probability $0.1$.
For each query $q \in Q^{\ast}$, we retrieve its top $10$ most similar items in $U$ 
and include them in the interest stream $I$ at
$q$'s timestamp $t_q$,
as well as at their original arrival times in $U$.
Table~\ref{table:dynp_stats} summarizes the item and interest stream statistics.
We compute popularity scores at the current time according to Definition~\ref{def:popScore}
with $\alpha=0.95$.

\begin{table*}[hbt]
\begin{center}
\begin{tabular}{|c|c|c|c|c|}
\hline
       		  & \multicolumn{2}{|c|}{Item stream}  & \multicolumn{2}{|c|}{Interest stream} \\ \hline
              & Num. items       & Num. ticks      & Num. items   & Num. ticks   \\ \hline
Reuters       & $540,\!882$      & 252             & $226,\!890$  & 95   \\   
Twitter       & $13,\!124,\!853$ & $2,\!000$ &    $4,\!267,\!518$ & $1,\!500$   \\  \hline

\end{tabular}
\vspace{1em}
\caption{\DPOP{} item and interest streams statistics.}
\label{table:dynp_stats}
\end{center}
\end{table*}

\begin{figure}[hbt]
    \centering
    	\subfloat[Reuters]{\includegraphics[scale=0.3]{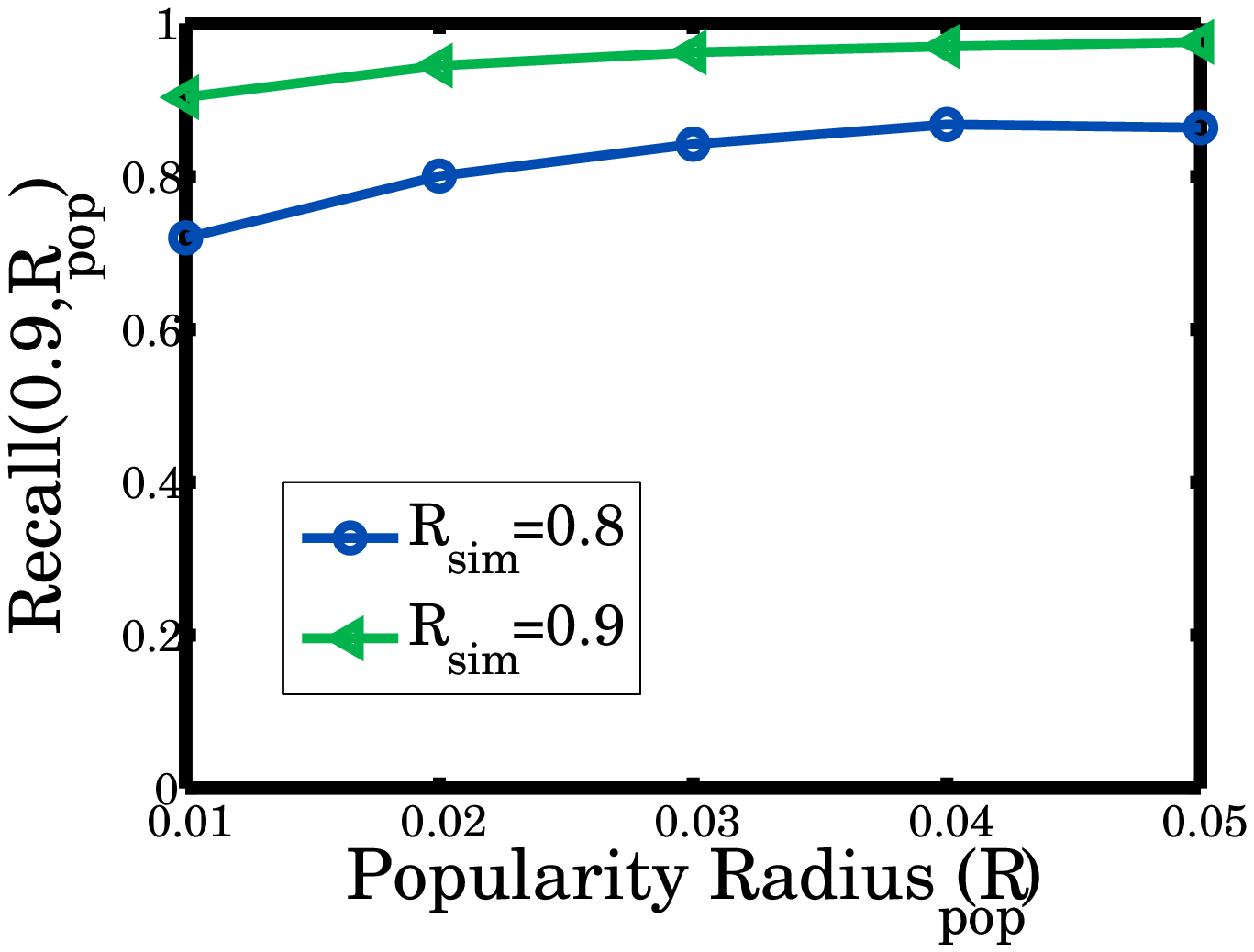}\label{subfig:reuters_dynp}}
		\subfloat[Twitter]{\includegraphics[scale=0.3]{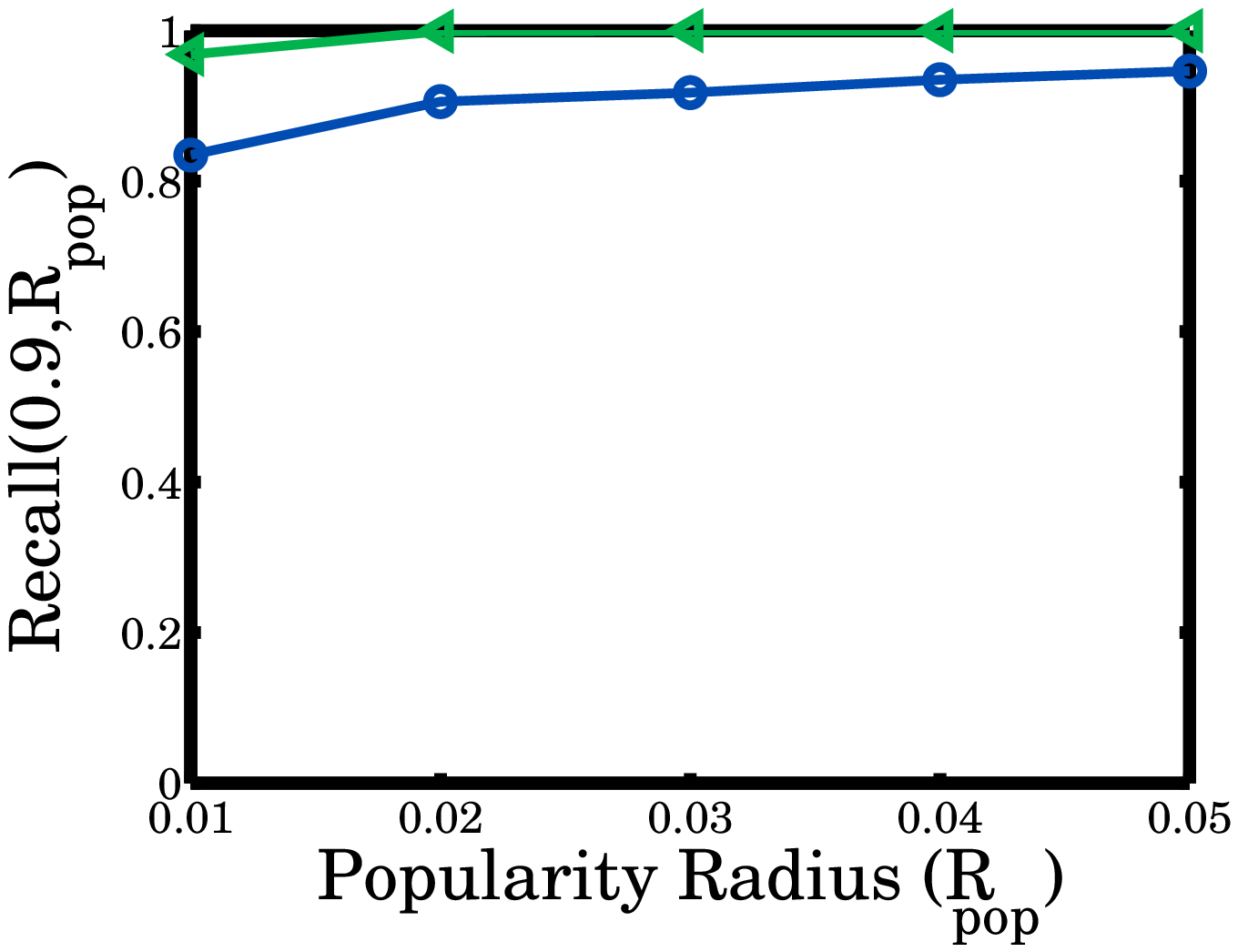}\label{subfig:twitter_dynp}}
	\vspace{1em}
    \caption{Recall by popularity radius of Stream-LSH when using \DPOP{} with the Smooth retention policy.}    
						\label{fig:dynpop_recall}
\end{figure}

Figure \ref{fig:dynpop_recall} depicts recall as a function of $R_{pop}$ 
for similarity radii $0.8$ and $0.9$.
For both datasets and similarity radii, the recall increases as the popularity radius increases.
\DPOP{} provides high recall when searching for the most popular items in the dataset:
For example, in the Reuter's dataset (Figure~\ref{fig:dynpop_recall}\subref{subfig:reuters_dynp}),
for $R_{sim}=0.8$ (blue curve) and $R_{pop} = 0.05$ (capturing the 3.5\% most popular items in the data set), 
the recall is $0.86$.
When increasing the similarity radius to $R_{sim}=0.9$ (green curve),
the recall increases and is $0.97$.
\DPOP{}'s recall is lower when we also search for less popular items:
In the Reuter's dataset, for $R_{sim}=0.8$ and $R_{pop} = 0.01$ (capturing the 24\% most popular items in the data set), 
the recall is $0.72$.
When increasing the similarity radius to $R_{sim}=0.9$, the recall is $0.9$.
Overall, \DPOP{} achieves good recall for popular items that are similar to the query
while also retrieving similar items that are less popular albeit with lower recall;
the latter is beneficial for applications such as query auto-completion~\cite{BarYossef11} 
and product recommendation~\cite{LongTailRec12}.

\section{Related Work}
\label{sec:RelatedWork}
Previous work on recommendation over streamed content~\cite{Newsjunkie2004,Das2007,Liu2010,Kompan2010,SCENE2011,SNR2014,EarlyBird}
focused on using temporal information for increasing the relevance of recommended items.
Stream recommendation algorithms extend techniques originally designed for static data,
such as content-based and collaborative-filtering~\cite{recommenderSurvey},
and apply them to streamed data
by taking into account the temporal characteristics of stream
generation and consumption within the algorithm internals. 
In the context of search,
many works extend 
ranking methods to consider temporal aspects of the data,
(see~\cite{TemporalIR} for a survey),
and quality features 
such as a social post's length or the 
author's influence~\cite{Twinder2012}.
However, these search and recommendation works do not tackle the challenge of bounding
the capacity of their underlying indexing data-structures. 
Rather, they assume an index of the entire stream with temporal information is given.
Our work is thus complementary to these efforts in the sense that we offer a
retention policy that may be used within their similarity search building block.

TI~\cite{TI2011} and LSII~\cite{LSII13} improve realtime indexing of stream data
using a policy that determines which items to index online and which to defer to a later batch indexing stage.
Both assume unbounded storage and are thus complementary to our work.
In addition, the TI focuses on highly popular queries, 
whereas we also address the tail of the popularity distribution.
LSII addresses the tail,
however, it assumes exact search while we focus on approximate search,
which is the common approach in similarity search~\cite{Gionis99}.

A few previous works have addressed bounding
the underlying index size in the context of stream processing~\cite{Petrovic2010,Sundaram2013,SSSJ16,Magdy16}.
Two papers~\cite{Petrovic2010,Sundaram2013} have focused on \emph{first story detection},
which detects new stories that were not previously seen.
Both use LSH as we do.
Petrovi\'{c} et al.~\cite{Petrovic2010}
maintain buckets of similar stories,
which are used in realtime for detecting new stories using similarity search.
In order to bound the index, they define
a limit on the number of stories kept within a bucket,
and eliminate the oldest stories when the limit is reached.
We call this retention policy Bucket.
Sundaram et al.~\cite{Sundaram2013}'s primary goal is to parallelize LSH,
in order to support high-throughput data streaming.
They bound the index size using a retention policy we call Threshold,
by eliminating the oldest items when the entire index exceeds a given space limit.
Both papers focus on the first story detection application,
while our work focuses on the similarity search primitive.
Their retention policies are well-suited for first story detection, 
where the index is searched in order to determine whether any recent matching result exists 
(indicating that the story is not the first), 
and are less adequate for similarity search, 
where multiple relevant results to an arbitrary query are targeted.
We evaluate their retention policies in our Stream-LSH algorithm and
find that our Smooth policy provides much
better results in our context.

Morales and Gionis propose \emph{streaming similarity self-join} (SSSJ)~\cite{SSSJ16},
a primitive that finds pairs of similar items within an unbounded data stream.
Similarly to us, SSSJ needs to bound its underlying search index.
Our work differs however in several aspects:
First, we study a different search primitive, namely, similarity search,
which searches for items similar to an arbitrary input query rather than retrieving pairs of similar items
from the stream.
Second, SSSJ only retrieves items that are not older than a given age limit.
It thus bounds the index using a variant of Threshold.
In contrast, we do not assume that an age limit on all queries is known a priory.
In this context, we propose Smooth, which better fits our setting as we show in our evaluation.
Third, we tackle approximate similarity search 
whereas SSSJ searches for an exact set of similar pairs.

Magdy et al.~\cite{Magdy16} propose a search solution over stream data with bounded storage,
which increases the recall of tail queries.
Their work differs from ours in the retrieval model,
more specifically, they assume
the ranking function is static and query-independent, e.g., ranking items by their age. 
Each item's score is known a priori for all queries,
and can be used to decide at indexing time which items to retain in the index.
This approach is less suitable to similarity search, 
where scores are query-dependent and only known at runtime.

In addition, we note that the aforementioned works on bounded-index stream processing~\cite{Petrovic2010,Sundaram2013,SSSJ16,Magdy16}
do not take into account quality and dynamic popularity as we do.

Several papers have focused on improving 
the space complexity of LSH via alternative search algorithms~\cite{Gan2012,Tao2010,SRS2014},
via decreasing the number of tables used at the cost of executing more queries~\cite{Entropy06}, 
or by searching more buckets~\cite{Multi-probe07}.
Unlike Stream-LSH, these works consider static (finite) data rather than a stream.

\section{Conclusions and Future Work}
\label{sec:conc}
We introduced the problem of similarity search over endless data-streams,
which faces the challenge of indexing unbounded data.
We proposed Stream-LSH, 
an SSDS algorithm that uses a retention policy to bound the index size.
We showed that our Smooth retention policy increases recall of similar items
compared to methods proposed by prior art.
In addition, our Stream-LSH indexing procedure is quality-sensitive,
and is extensible to dynamically retain items according to their popularity.

While our work focuses on similarity search,
our approach may prove useful in future work, for addressing space constraints 
in other stream-based search and recommendation primitives.

\balance 

\bibliographystyle{abbrv}
\begin{scriptsize}
\bibliography{stream-lsh}  

\begin{thebibliography}{10}

\bibitem{Lucene}
Lucene.
\newblock \url{http://lucene.apache.org/core/}.

\bibitem{Reuters}
Reuters rcv1.
\newblock \url{http://www.daviddlewis.com/resources/testcollections/rcv1/}.

\bibitem{TarsosLSH}
Tarsos-lsh.
\newblock \url{https://github.com/JorenSix/TarsosLSH}.

\bibitem{zephoria}
The top 20 valuable facebook statistics.
\newblock \url{https://zephoria.com/top-15-valuable-facebook-statistics/}.

\bibitem{TwitterNas}
Twitter nasdaq.
\newblock
  \url{http://followthehashtag.com/datasets/nasdaq-100-companies-free-twitter-dataset/}.

\bibitem{recommenderSurvey}
G.~Adomavicius and A.~Tuzhilin.
\newblock Toward the next generation of recommender systems: A survey of the
  state-of-the-art and possible extensions.
\newblock {\em IEEE transactions on knowledge and data engineering},
  17(6):734--749, 2005.

\bibitem{Agichtein2008}
E.~Agichtein, C.~Castillo, D.~Donato, A.~Gionis, and G.~Mishne.
\newblock Finding high-quality content in social media.
\newblock WSDM '08, pages 183--194, 2008.

\bibitem{Andoni09NNS}
A.~Andoni.
\newblock {\em Nearest Neighbor Search: the Old, the New, and the Impossible}.
\newblock PhD thesis, Massachusetts Institute of Technology, 2009.

\bibitem{BarYossef11}
Z.~Bar-Yossef and N.~Kraus.
\newblock Context-sensitive query auto-completion.
\newblock WWW '11, pages 107--116, 2011.

\bibitem{Becker11}
H.~Becker, M.~Naaman, and L.~Gravano.
\newblock Selecting quality twitter content for events.
\newblock ICWSM11, 2011.

\bibitem{FlickrTemporalSearch16}
F.~R. Bentley, J.~J. Kaye, D.~A. Shamma, and J.~A. Guerra-Gomez.
\newblock The 32 days of christmas: Understanding temporal intent in image
  search queries.
\newblock CHI '16, pages 5710--5714, 2016.

\bibitem{EarlyBird}
M.~Busch, K.~Gade, B.~Larson, P.~Lok, S.~Luckenbill, and J.~Lin.
\newblock Earlybird: Real-time search at twitter.
\newblock ICDE '12, pages 1360--1369, 2012.

\bibitem{Charikar02}
M.~S. Charikar.
\newblock Similarity estimation techniques from rounding algorithms.
\newblock In {\em STOC '02}, pages 380--388, 2002.

\bibitem{TI2011}
C.~Chen, F.~Li, B.~C. Ooi, and S.~Wu.
\newblock Ti: An efficient indexing mechanism for real-time search on tweets.
\newblock SIGMOD '11, pages 649--660, 2011.

\bibitem{Chen2011}
J.~Chen, R.~Nairn, and E.~Chi.
\newblock Speak little and well: Recommending conversations in online social
  streams.
\newblock CHI '11, pages 217--226, 2011.

\bibitem{Chierichetti12}
F.~Chierichetti and R.~Kumar.
\newblock {LSH}-preserving functions and their applications.
\newblock In {\em SODA '12}, pages 1078--1094, 2012.

\bibitem{unicorn2013}
M.~Curtiss, I.~Becker, T.~Bosman, S.~Doroshenko, L.~Grijincu, T.~Jackson,
  S.~Kunnatur, S.~Lassen, P.~Pronin, S.~Sankar, G.~Shen, G.~Woss, C.~Yang, and
  N.~Zhang.
\newblock Unicorn: A system for searching the social graph.
\newblock {\em Proc. VLDB Endow.}, pages 1150--1161, 2013.

\bibitem{Das2007}
A.~S. Das, M.~Datar, A.~Garg, and S.~Rajaram.
\newblock Google news personalization: Scalable online collaborative filtering.
\newblock WWW '07, pages 271--280, 2007.

\bibitem{Diaz2009}
F.~Diaz.
\newblock Integration of news content into web results.
\newblock WSDM '09, pages 182--191, 2009.

\bibitem{Newsjunkie2004}
E.~Gabrilovich, S.~Dumais, and E.~Horvitz.
\newblock Newsjunkie: Providing personalized newsfeeds via analysis of
  information novelty.
\newblock WWW '04, pages 482--490, 2004.

\bibitem{Gan2012}
J.~Gan, J.~Feng, Q.~Fang, and W.~Ng.
\newblock Locality-sensitive hashing scheme based on dynamic collision
  counting.
\newblock SIGMOD '12, pages 541--552, 2012.

\bibitem{Gionis99}
A.~Gionis, P.~Indyk, and R.~Motwani.
\newblock Similarity search in high dimensions via hashing.
\newblock In {\em VLDB '99}, pages 518--529, 1999.

\bibitem{Guy:2012}
I.~Guy, T.~Steier, M.~Barnea, I.~Ronen, and T.~Daniel.
\newblock Swimming against the streamz: Search and analytics over the
  enterprise activity stream.
\newblock CIKM '12, pages 1587--1591, 2012.

\bibitem{LSH_Indyk1998}
P.~Indyk and R.~Motwani.
\newblock Approximate nearest neighbors: Towards removing the curse of
  dimensionality.
\newblock STOC '98, pages 604--613, 1998.

\bibitem{TemporalIR}
N.~Kanhabua, R.~Blanco, and K.~N{\o}rv{\aa}g.
\newblock Temporal information retrieval.
\newblock {\em Foundations and Trends in Information Retrieval}, pages 91--208,
  2015.

\bibitem{Kompan2010}
M.~Kompan and M.~Bielikova.
\newblock Content-based news recommendation.
\newblock In {\em E-Commerce and Web Technologies}, pages 61--72. 2010.

\bibitem{WhatIsTwitter10}
H.~Kwak, C.~Lee, H.~Park, and S.~Moon.
\newblock What is twitter, a social network or a news media?
\newblock WWW '10, pages 591--600, 2010.

\bibitem{MiningMassive}
J.~Leskovec, A.~Rajaraman, and J.~D. Ullman.
\newblock {\em Mining of Massive Datasets, 2nd Ed}.
\newblock Cambridge University Press, 2014.

\bibitem{SCENE2011}
L.~Li, D.~Wang, T.~Li, D.~Knox, and B.~Padmanabhan.
\newblock Scene: A scalable two-stage personalized news recommendation system.
\newblock SIGIR '11, pages 125--134, 2011.

\bibitem{Liu2010}
J.~Liu, P.~Dolan, and E.~R. Pedersen.
\newblock Personalized news recommendation based on click behavior.
\newblock IUI '10, pages 31--40, 2010.

\bibitem{SNR2014}
M.~Lu, Z.~Qin, Y.~Cao, Z.~Liu, and M.~Wang.
\newblock Scalable news recommendation using multi-dimensional similarity and
  jaccard-kmeans clustering.
\newblock {\em J. Syst. Softw.}, pages 242--251, 2014.

\bibitem{Multi-probe07}
Q.~Lv, W.~Josephson, Z.~Wang, M.~Charikar, and K.~Li.
\newblock Multi-probe lsh: Efficient indexing for high-dimensional similarity
  search.
\newblock In {\em VLDB '07}, pages 950--961, 2007.

\bibitem{Magdy16}
A.~Magdy, R.~Alghamdi, and M.~F. Mokbel.
\newblock On main-memory flushing in microblogs data management systems.
\newblock ICDE '16, pages 445--456, 2016.

\bibitem{SSSJ16}
G.~D.~F. Morales and A.~Gionis.
\newblock Streaming similarity self-join.
\newblock {\em {PVLDB}}, 9(10):792--803, 2016.

\bibitem{Entropy06}
R.~Panigrahy.
\newblock Entropy based nearest neighbor search in high dimensions.
\newblock In {\em SODA '06}, pages 1186--1195, 2006.

\bibitem{Petrovic2010}
S.~Petrovi\'{c}, M.~Osborne, and V.~Lavrenko.
\newblock Streaming first story detection with application to twitter.
\newblock HLT '10, pages 181--189, 2010.

\bibitem{NNSTutorial}
M.~Slaney and M.~Casey.
\newblock Locality-sensitive hashing for finding nearest neighbors.
\newblock {\em Signal Processing Magazine, IEEE}, pages 128--131, 2008.

\bibitem{SRS2014}
Y.~Sun, W.~Wang, J.~Qin, Y.~Zhang, and X.~Lin.
\newblock Srs: Solving c-approximate nearest neighbor queries in high
  dimensional euclidean space with a tiny index.
\newblock {\em Proc. VLDB Endow.}, pages 1--12, 2014.

\bibitem{Sundaram2013}
N.~Sundaram, A.~Turmukhametova, N.~Satish, T.~Mostak, P.~Indyk, S.~Madden, and
  P.~Dubey.
\newblock Streaming similarity search over one billion tweets using parallel
  locality-sensitive hashing.
\newblock {\em Proc. VLDB Endow.}, 6(14):1930--1941, Sept. 2013.

\bibitem{Twinder2012}
K.~Tao, F.~Abel, C.~Hauff, and G.-J. Houben.
\newblock Twinder: A search engine for twitter streams.
\newblock ICWE'12, pages 153--168, 2012.

\bibitem{Tao2010}
Y.~Tao, K.~Yi, C.~Sheng, and P.~Kalnis.
\newblock Efficient and accurate nearest neighbor and closest pair search in
  high-dimensional space.
\newblock {\em ACM Trans. Database Syst.}, pages 20:1--20:46, 2010.

\bibitem{TwitterSearch11}
J.~Teevan, D.~Ramage, and M.~R. Morris.
\newblock \#twittersearch: A comparison of microblog search and web search.
\newblock WSDM '11, pages 35--44, 2011.

\bibitem{Weber1998}
R.~Weber, H.-J. Schek, and S.~Blott.
\newblock A quantitative analysis and performance study for similarity-search
  methods in high-dimensional spaces.
\newblock VLDB '98, pages 194--205, 1998.

\bibitem{LSII13}
L.~Wu, W.~Lin, X.~Xiao, and Y.~Xu.
\newblock {LSII:} an indexing structure for exact real-time search on
  microblogs.
\newblock ICDE '12, pages 482--493, 2013.

\bibitem{TweeterSNAP}
J.~Yang and J.~Leskovec.
\newblock Patterns of temporal variation in online media.
\newblock WSDM '11, pages 177--186, 2011.

\bibitem{LongTailRec12}
H.~Yin, B.~Cui, J.~Li, J.~Yao, and C.~Chen.
\newblock Challenging the long tail recommendation.
\newblock {\em Proc. VLDB Endow.}, pages 896--907, 2012.

\end{thebibliography}
\end{scriptsize}

\end{document}